\PassOptionsToPackage{warn}{textcomp}
\documentclass[nonacm]{acmart}

\usepackage{booktabs} 
\usepackage{epsfig,endnotes, graphicx}  
\usepackage{multirow}  
\usepackage{url}
\usepackage{balance}
\usepackage{times}
\usepackage{subcaption}
\usepackage{color}
\usepackage{anyfontsize}
\usepackage{amsmath}
\usepackage{amssymb}
\usepackage{hyperref}
\usepackage[ruled, vlined, commentsnumbered]{algorithm2e}
\usepackage{algpseudocode,algorithmicx}
\usepackage{caption}

\usepackage{pifont}
\usepackage{tablefootnote}

\setcopyright{none}

\hypersetup{draft}

\begin{document}
\title{RF Soil Moisture Sensing via Radar Backscatter Tags}
 \author{Colleen Josephson}
  \affiliation{
    \institution{Stanford University}
 }
 \email{cajoseph@stanford.edu}

\author{Bradley Barnhart}
 \affiliation{
   \institution{Stanford University}
 }
 \email{bbarnhart@stanford.edu}

\author{Sachin Katti}
 \affiliation{
   \institution{Stanford University}
 }
 \email{skatti@cs.stanford.edu}

\author{Keith Winstein}
 \affiliation{
   \institution{Stanford University}
 }
 \email{keithw@cs.stanford.edu}

\author{Ranveer Chandra}
 \affiliation{
   \institution{Microsoft}
 }
 \email{ranveer@microsoft.com}
\renewcommand{\shortauthors}{C. Josephson, B. Barnhart, S. Katti, K. Winstein, R. Chandra}

\begin{abstract}

  We present a sensing system that determines soil moisture
  via RF using backscatter tags paired with a commodity ultra-wideband
  RF transceiver. Despite decades of research confirming the benefits,
  soil moisture sensors are still not widely adopted on working farms
  for three key reasons: the high cost of sensors, the difficulty of
  deploying and maintaining these sensors, and the lack of reliable
  internet access in rural areas. We seek to address these obstacles
  by designing a low-cost soil moisture sensing
  system that uses a hybrid approach of pairing completely wireless backscatter tags with a mobile reader.

  We designed and built two backscatter tag prototypes and tested our
  system both in laboratory and \emph{in situ} at an organic farm
  field. Our backscatter tags have a projected battery lifetime of up
  to 15 years on $4\times$AA batteries, and can operate at a depth of at least 30cm and up to 75cm. We achieve an average accuracy within 0.01-0.03$cm^3/cm^3$ of the ground truth with a 90th percentile of $0.034cm^3/cm^3$, which is comparable to
  state-of-the-art commercial soil sensors, at an order of magnitude lower cost.
\end{abstract}




\maketitle

\setlength{\textfloatsep}{5pt}

\begin{figure}
 \centering \includegraphics[scale=0.25]{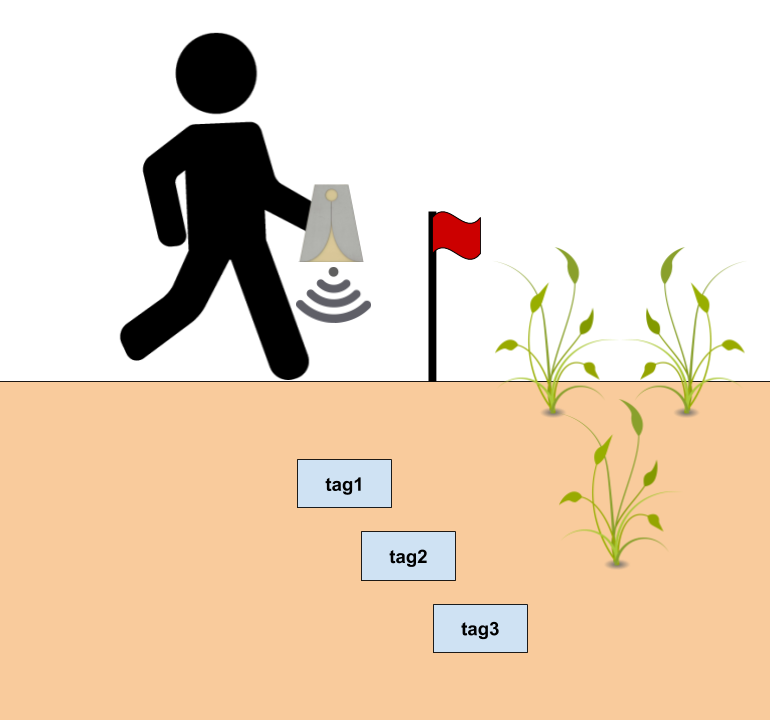}\\
 \caption{Our system uses an affordable portable radar paired with
   inexpensive underground backscatter tags to measure soil moisture}
  \label{figure:phone}
\end{figure}

\section{Introduction}
Agriculture is the single largest pressure on the world's sources of
fresh water--- 69\% of the global fresh water supply is used for
agriculture~\cite{water}. Paired with the fact that the global
population is projected to exceed 9 billion by 2050~\cite{population}
with most of that growth coming from developing nations in Africa and
Asia, conservation of fresh water and sufficient food production are
key concerns that need to be addressed for future generations. Soil
moisture is the most important measurement for ensuring the
maximization of crop yield without water waste.

Multiple studies show that soil moisture sensors lead to a water
savings of at least 15\%\cite{watersavings}, and in some cases more
than 50\%, while maintaining crop yields or even increasing them up to
26\%~\cite{Zotarelli2009}. Yet soil sensors are still not widely
deployed on working farms despite decades of research confirming the
benefits.  Fewer than 10\% of irrigated crops in the United States use moisture
sensors~\cite{USDAirrigation}, and that number is even lower in
developing nations. The lack of widespread adoption can be attributed
to three key challenges: 1.) high sensor cost 2.)  difficulty of
deploying and maintaining the sensors and 3.) difficulty collecting
and processing the sensor data.

The average commercial soil moisture sensor costs more than \$100,
which does not include a power source or data logger to record and/or
transmit the measurement samples. Since soil moisture is not uniform
across a field, multiple sensors are needed to accurately measure
moisture for irrigation purposes. The average farm in the United
States is 444 acres~\cite{farmsize}. For a farm of that size, the
conservative cost of deploying the recommended density of 20
sensors~\cite{sensorDensity} per acre would be more than a million
dollars. Even for sparse deployments, benefits exceed the costs
only a third of the time~\cite{USDAprofits}. This makes it difficult
for farmers in even the wealthiest nations to justify investing in
moisture sensors. Consequently, soil moisture sensing is currently
infeasible for smallholder farmers in developing nations, which is
where the most of the food and water insecurity will be.

In addition to cost barriers, current sensors are not simple to deploy
and maintain. Very few companies offer a product that includes the
sensor, logger and power source ready for immediate use. Therefore
some amount of setup labor is required for each sensor. Though the
sensor is waterproof, the data logger (e.g., an Arduino) may need to be
waterproofed and powered separately. The sensor probe also needs to be buried. To
supply power, many opt to attach solar panels to a battery pack, which
requires mounting the panels to a wooden or metal post. These
laborious processes needs to be repeated for every sensor node on the
farm, and again each time the field is tilled/cultivated. Furthermore, excess cables and bulky battery boxes make the
system prone to entanglement in farm equipment and tools. This all
adds up to a significant amount of manual labor to deploy and maintain the
sensor network.


Finally, data needs to be collected from loggers. For a large farm,
the most practical method is wireless collection. Installing WiFi or
cellular communication modules on every sensor is costly and
exacerbates power issues. Extending wireless coverage to a large farm
is not simple; cellular coverage in rural areas tends to be poor. A
number of recent works have considered the issue of networking sensors
in rural environments. For example,~\cite{farmbeats} uses TV
whitespace technology to provide a wireless gateway from the field to
the Internet. LoRaWAN, Sigfox and NB-IoT are all low-power wide-area networks (LPWANs) that target large scale IoT deployments~\cite{RazaKS16}.

In contrast to the networked sensor model used on farms, geophysicists
and remote sensing experts use a centralized approach. They have been
using ground penetrating radar (GPR) instead of wired sensors to
measure soil moisture for years. GPR has the advantage that it can
measure soil moisture completely wirelessly eliminating the need for
sensor probes, solar panels and data loggers. The signal strength and
propagation speed of an RF wave is impacted by the media it travels
though. RF travels 2-6 times more slowly in soil than air~\cite{Jol2008},
and the speed and signal strength decrease as moisture content
increases. Radars allow us to very accurately measure these changes RF
waves.

The drawback is that the radars used in these studies are either
deployed in satellites~\cite{Fares2013} whose data do not provide the necessary resolution, or use terrestrial radars that require contact
(or very close proximity) with the ground~\cite{Shamir2018}. These
terrestrial GPRs are also bulky, most being at least the size of a
lawnmower. Furthermore, the depth and accuracy achievable using a
terrestrial GPR alone is limited. This makes traditional GPR soil
moisture measurement techniques impractical for agriculture.

To address these concerns, we propose a hybrid approach. Instead of
using radar alone, we pair the radar with completely wireless
underground backscatter tags. Unlike traditional backscatter tags,
these tags do not have any additional sensors attached to them whose
measurements need to be communicated. Instead they merely provide a
known reference point in the ground and increase the strength of the
signal returning to the radar. This allows us to measure soil moisture
with RF using a significantly cheaper and more portable radar than
traditional terrestrial GPRs. In the future this radar reader could
even be integrated with farm equipment, drone or mobile phone.

This two-part system allows us to implement a low-maintenance and
low-cost soil moisture sensing system that does not require cellular or other wireless connectivity. Backscatter tags are more simple to install than wired sensors, and do not require additional power or network infrastructure. The tags enable the radar reader to be mobile, which makes collecting measurements much easier and less destructive than using a traditional rolling GPR. The components are also lower cost than traditional moisture measurement systems. Mass-produced weather-poof backscatter tags such as~\cite{greenlee} are between \$5-10, and consumer-grade UWB radars are \$400-2500.

We designed two
prototype backscatter tags: one active and the other
semi-passive. Both measured soil moisture with an average error of
0.01-0.02$cm^3/cm^3$, a 90th percentile of $0.034cm^3/cm^3$, and a maximum error of at most 0.055$cm^3/cm^3$,
which is comparable to the accuracy of commercial soil
sensors~\cite{Datta2018}. The active tag measures soil moisture
accurately to saturation with a projected battery life of 3-4 years (3 months without duty cycling) on $4\times$AA batteries, where the semi-passive tag is accurate within ranges typically
seen in agriculture and has a projected battery life of more than 15
years. We also show that the system can deployed at depths of 30cm or
more.

\section{Background}

The volumetric water content (VWC) of soil, often represented as
$\Theta$, is defined as the ratio of the volume of the water in the
soil to the volume of the soil plus water:

\begin{equation}
  \Theta = \frac{V_{water}}{V_{wet soil}}
\end{equation}
  
It is measured in units of $cm^3/cm^3$. The most accurate way to
measure VWC is by taking a soil sample of a known volume, weighing it,
drying it in an oven for 24+ hours, and then re-weighing
it~\cite{Noborio2001}. This process is time-consuming and requires
physical removal of soil at the depth you wish to measure, which makes
it impractical for irrigation purposes. Instead, commercial sensors
approximate VWC by measuring properties that are closely
correlated. One such property is the permittivity, $\varepsilon$,
which increases with the VWC of soil.

Recall that permittivity is the ability of a substance to hold an
electrical charge. It is often treated as a complex number:
\begin{equation}
  \varepsilon = \varepsilon' + j\varepsilon ''
\end{equation}
where $\varepsilon '$ is the real component and $\varepsilon''$ is the
complex.

The \emph{dielectric permittivity constant} (also known as relative
permittivity), $\varepsilon_r$, is the ratio of the permittivity of
the substance to the permittivity of free space, $\varepsilon_0$:

\begin{equation}
  \varepsilon_r = \varepsilon/\varepsilon_0
\end{equation}

There are two main types of sensors that measure permittivity to
approximate soil moisture: capacitive and time domain reflectometry
(TDR). 

\subsection{Capacitive sensors}
Capacitive sensors measure the charge time of a capacitor,
which is a roughly linear function of
$\varepsilon$~\cite{sensorOverview}. The resistance of the soil is
also used to measure moisture, since as moisture increases resistance
decreases. However, every measurement degrades a resistive sensor via
electrolysis. Capacitive sensors are less prone to corrosion than
resistive sensors and are more accurate. For this reason,
commercial-grade soil moisture sensors are usually capacitive instead
of resistive\footnote{The SparkFun soil moisture sensor retailing for
  \$6.95 is resistive and known for corroding quickly, so it is not
  used on farms}. In this work we compare the accuracy of our system
against a Teros 12 capacitive sensor, which retails for \$250.

\subsection{TDR sensors}
Time domain reflectometry (TDR) is another common method for measuring
soil moisture. It measures the propagation time of EM waves by sending
a pulse down a cable and into the soil probe and measures how long it
takes for the signal to return. This time, represented as $\tau$, is
often called \emph{time of flight} (ToF) in the context of wireless
transmissions. $\tau$ is used to approximate the \emph{apparent
  dielectric constant} $K_a$, which is a function of
$\varepsilon_r', \varepsilon_r'', \varepsilon_0$ and electrical
conductivity (EC) $\sigma$:

\begin{equation}
  K_a = \frac{\varepsilon_r'}{2}\Bigg[\sqrt{1+\bigg(\frac{\varepsilon_r'' + \frac{\sigma}{2\pi f\varepsilon_0}}{\varepsilon_r'}\bigg)^2}+1\Bigg]  
\end{equation}

At high frequencies, $\epsilon_r$ is dominated by the real part
$\epsilon_r'$, so 

\begin{equation}
  K_a \approx \varepsilon_r'
\end{equation}

The velocity of a wave in a media is

\begin{equation}
  v = c \left(\frac{\mu\varepsilon_r'}{2} \left[1+\sqrt{1+ \left(\tfrac{\sigma}{\omega}\right)^2}\right] \right)^{-1/2}
\end{equation}

where $c$ is the speed of light in free space, $\mu$ is the relative
magnetic permeability of the material, $\sigma$ is the conductivity (EC)
and $\omega$ is the angular frequency of the wave.

In soil, the magnetic permeability is very close to
1 H/m~\cite{patitz1995measurement} and the conductivity is typically less
than 0.3 S/m ~\cite{idealEC} for any soil. Therefore, when $\omega$ is sufficiently large the velocity simplifies to

\begin{equation}
  v = c/\sqrt{\varepsilon_r'}
\end{equation}

If we know the distance $d$ that the wave travels through soil and the
ToF $\tau$, then $v = d/\tau$ and 

\begin{equation}
K_a \approx \left(\frac{c\tau}{d}\right)^2
\end{equation}

Soil moisture $\Theta$ can be related to $K_a$ using formulas that depend on the soil type. One such formula is the Topp Equation ~\cite{Topp1980}, which is applicable to typical soils\footnote{A typical soil is 50\% solids (45-49\% minerals, 1-6\% organic matter) and 50\% water/air~\cite{typicalSoil}}:

\begin{equation}
  \Theta = 4.3\times 10^{-6}K_a^3-5.5\times10^{-4}K_a^2+2.92\times 10^{-2}K_a-5.3\times 10^{-2}
\end{equation}

Substituting,

\begin{equation}
  \Theta \approx 4.3\times 10^{-6} (\tfrac{c\tau}{d})^6-5.5\times10^{-4} (\tfrac{c\tau}{d})^4+2.92\times 10^{-2} (\tfrac{c\tau}{d})^2-5.3\times 10^{-2}
\end{equation}

Therefore, if we know the distance $d$ that an RF wave travels, and we
can accurately measure ToF, then we are able to approximate $\Theta$.

In TDR soil sensors, RF travels along a waveguide of a known
distance. They generate wideband signals (100Mhz-3Ghz for agricultural
grade) to ensure sufficient time resolution. Like capacitive sensors,
the probes must maintain good contact with the soil for accurate
measurements. The cost of a TDR sensor is about \$1000. Next we
discuss how soil moisture can be measured wirelessly without a
waveguide using radar technology.

\subsection{RADAR and GPR}

RADAR (RAdio Detection And Ranging), colloquialized as radar, is a
well-developed technology with a history dating to before World War
II. Radars use the principle of \emph{RF backscatter}, phenomenon of
RF bouncing off reflectors back towards the originating transmitter. A
radar transmitter sends out a known waveform, and the receiver
correlates the incoming RF with that known waveform. The received
samples are used to determine Time of Flight and angle of arrival
information, which can then be used to calculate the distance of
objects and their speed. This technique is conceptually similar to
echolocation. Radar was originally used primarily in military
contexts, but nowadays it has a diverse set of applications such as
predicting weather patterns, enforcing road speed, assisting
autonomous vehicles with navigation~\cite{Ward2016}, and even
monitoring human breathing~\cite{Li2016}.

There are two primary types of radar waveforms: continuous wave and
pulsed~\cite{Richards2010}. Continuous wave radars are transmitting
and receiving at all times, while pulsed radars periodically transmit a
short-duration pulse and listen for the reflections to come back.

Pulsed radars can easily determine the distance of a target by
measuring the time that elapses between pulse transmission and the
return reflection. They can also measure target speed. When the pulse
width of the radar is very short, it is known as an impulse radar or
Ultra Wideband (UWB) radar~\cite{hussain1998ultra}. Commodity UWB radars were enabled by a key circuits
discovery in 1994: the single-shot transient
digitizer~\cite{Azevedo1997}. This device is capable of high speed,
high accuracy digitization of very short pulses of energy (<
5ns). This allowed for the construction of a significantly cheaper and
smaller radar transceiver that digitizes incoming RF and correlates it
with the transmitted pulse samples.

Impulse radar needs to be wideband because of Fourier duality: pulses
that are short in the time domain require wide bandwidth in the
frequency domain. The bandwidth of a UWB radar typically ranges
between 2 and 8 Ghz. Furthermore, the transmit power of UWB radars is
usually regulated to be very low to avoid causing interference to
other users on the same spectrum. This also makes UWB radar very
difficult to detect, as the transmitted waveform looks like white
noise. UWB radars are also popular because the low transmission power
ensures that the signal is harmless to living organisms.

Ground penetrating radars (GPR) use low frequencies that can penetrate
under the earth to do underground imaging. In addition to imaging,
these radars can also be used to calculate soil moisture using ToF
and/or signal strength. GPRs are usually wideband and cost thousands
of dollars. Generally the equipment is large (the size of a lawnmower
or larger) and needs to be dragged across the surface of the
soil. This is a labor intensive process which may not always be
possible in dense crops. Non-contact GPRs exist that can be attached
to drones, but these have lower resolution and can only reliably
measure to a depth of about 10cm~\cite{wu2019new}.

In the next section we discuss how pairing a radar with an RFID-like
underground backscatter tags allows us to measure soil moisture with
consumer-grade radars that are considerably less expensive and more
portable than traditional GPRs.

\section{Design}

\begin{figure}
 \centering \includegraphics[scale=0.25]{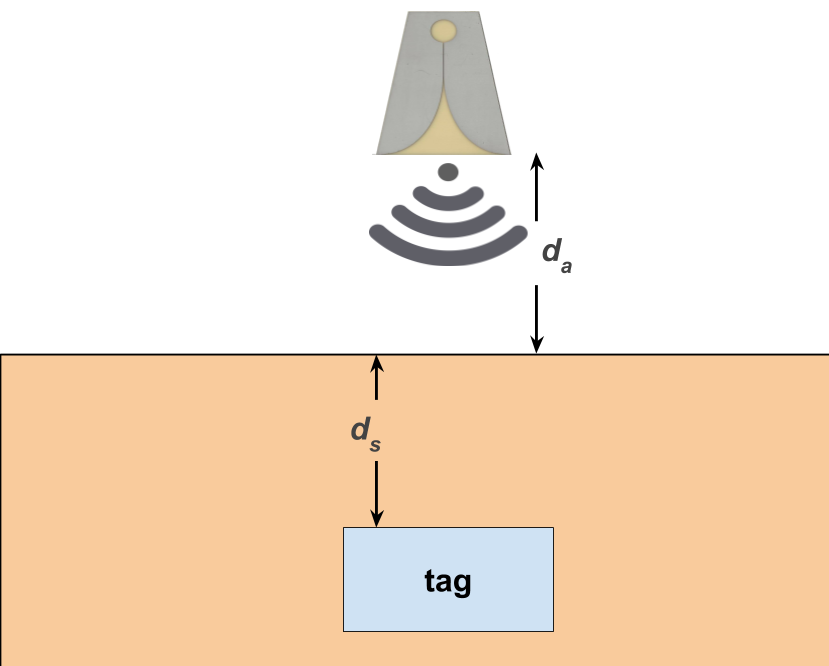}\\
 \caption{The tag is buried under soil at a known depth
   $d_s$. The radar can measure the ToF/distance between itself and
   the surface of the ground, $d_a$. Using these values, we can
   calculate $\Delta \tau$, the amount ToF increases due to traveling
   through soil instead of air}
  \label{figure:tagModel}
\end{figure}

Our key design insight is that we can make soil moisture
sensing orders of magnitude more affordable by using a system model
similar to RFID: cheap tags paired with a more expensive reader. Using
a GPR as the reader is possible, but these devices are often
 not small enough to be hand-held or mounted on a drone. Instead, we look to consumer-grade
radars. Consumer-grade radar systems are becoming increasingly
affordable and accessible, which introduces exciting new sensing
possibilities. There are multiple radar devices on the market as cheap as \$50-400, and modern smartphones have even started integrating
them~\cite{soli}. In addition to being affordable, these radars tend to
be lightweight and portable/handheld.

Consumer-grade UWB radars have the ability to accurately measure the
ToF of reflected signals, since their high bandwidth corresponds to a
time resolution of 0.15-0.5ns. For comparison, agricultural TDR
sensors range from 100Mhz to 3Ghz~\cite{Pelletier2012} which leads to
a resolution of at most 0.33ns. Therefore an appropriate selection of
UWB radar will allow us to measure ToF with the same accuracy as TDR
soil sensors.

Accurate measurement of ToF is not alone sufficient for measuring soil
moisture, though. A reference point buried a known distance beneath
the soil is required. One work used a large metallic object object
underground~\cite{Shamir2018}. We realized that versus a plain piece
of metal, using a grounded wideband directional antenna would
significantly increase the SNR of the signal returning to the
radar. This allows us to deploy the reference points deeper, and collect the measurements with inexpensive consumer-grade radars. These
radars also have a higher center frequency than GPRs, which allows
for smaller antennas and increased portability.

At mass production, these radar backscatter tags would cost similarly
to underground RFID tags that municipalities use for utility marking,
which cost \$5-10 in bulk. The cost to densely outfit a large
farm with tags and a few radars would be less then \$100,000, which is under a
tenth the cost of using traditional soil sensors. Like traditional sensors, the tags would be re-usable across growing seasons. Furthermore, if the tag is buried sufficiently deep it could even remain in the ground if the field is cultivated/tilled, which disturbs the top 15-30cm of soil~\cite{till}.

For a smaller farm, 1-2 acres, taking measurements by hand is
feasible. The reader could even be integrated with a mobile phone,
which would significantly facilitate adoption in developing
nations. For larger farms, taking readings by hand may not scalable,
so we assume that the reader would be mounted to a tractor, or
possibly an agricultural robot~\cite{Aroca2018} or
drone~\cite{farmbeats}, which are becoming increasingly popular.

In the following sections we expand on the design
details behind our system.

\subsection{Radar considerations}

A modern radar operates using \emph{frames}, which is the set of
samples from a single sweep across the radar's sensing area. For
example, a radar could sweep everything in front of it that lies
greater than 1 meter but less than 2 meters away. The sensing area is
typically adjustable. The shape of the sensing area depends on the
number and type of antennas. We use a pair (one TX, one RX) of
directional antennas, as the area of interest lies straight down.

For a pulsed radar, a frame typically has one complex sample per
\emph{range bin}. Each range bin corresponds to a range of possible
distances from the radar. For example, if the radar range resolution
is 5cm, the magnitude of the sample from the 10th range bin would
correspond to an object that is 45-50cm away from the radar. The
number and size of range bins depends on both the bandwidth and
sensing area. The wider the bandwidth, the smaller the range bins can
be. The larger the sensing area, the larger the range bins will
be. The size of the range bin determines the ToF resolution as well.

Three key specifications driving our choice of radar are the
bandwidth, frame rate. and center frequency.  As discussed earlier, any
UWB radar with a sufficiently wide bandwidth will provide suitable ToF
resolution. All of the radars we considered had a bandwidth of 3 or
more Ghz.

The frame rate determines how fast the radar can sweep the sensing
area. A faster frame rate means that measurements can be taken more
quickly and/or at a higher SNR. A number of radar settings impact the
achievable frame rate, including the sensing area and ADC/DAC
levels. All three of the radars we tested achieved a frame rate of at
least 200fps. For moisture levels typically seen on farms, this allows
us to take measurements within 10s for a tag buried at a depth
of 30cm.

\begin{figure}
 \centering \includegraphics[scale=0.5]{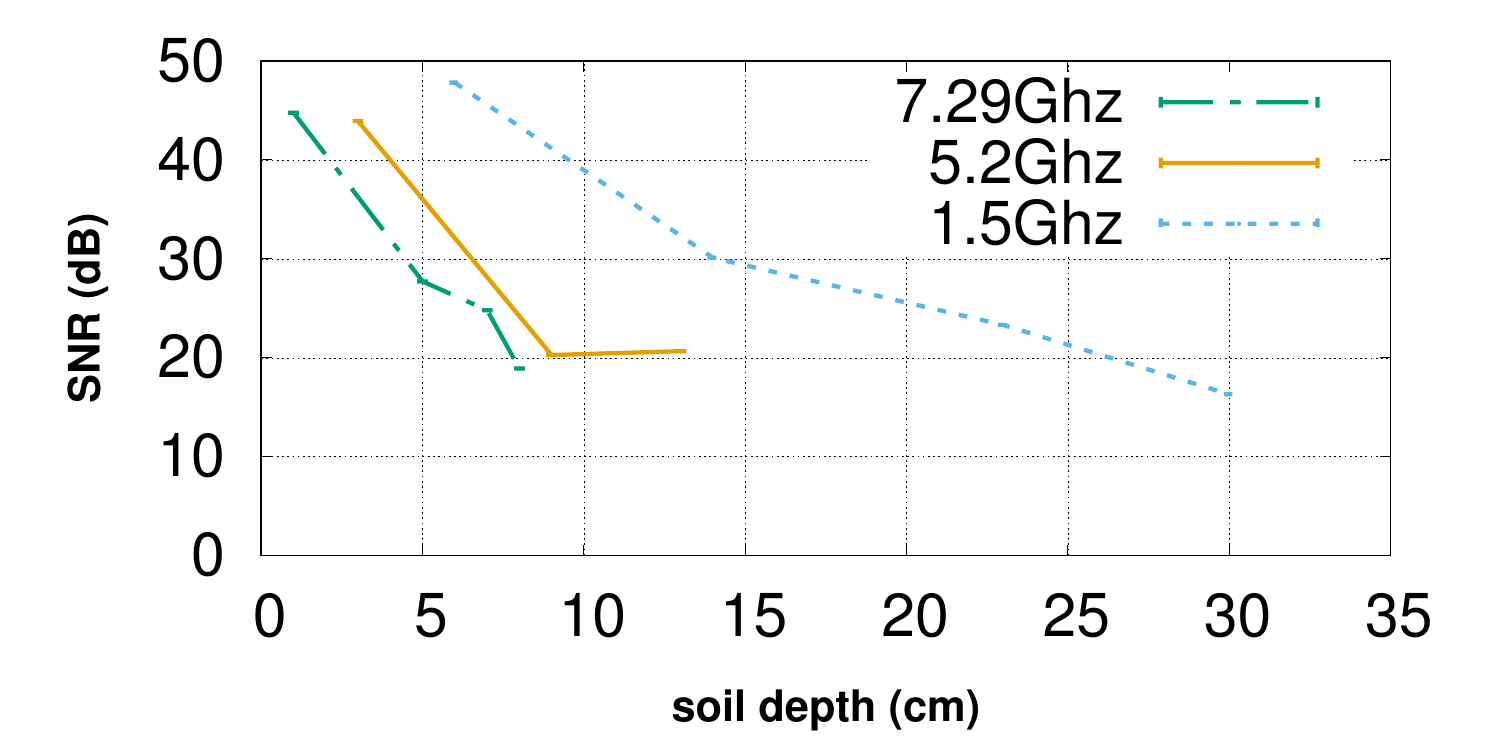}\\
 \caption{SNR vs soil depth for three radar center frequencies}
  \label{figure:radarSNR}
\end{figure}

The center frequency determines how deep beneath the ground a radar
can penetrate with a given transmission power. As the wavelength of
the RF decreases, the wave attenuates faster due to obstacles and
water (see Fig.~\ref{figure:radarSNR}). We ultimately selected the radar
centered at 1.5Ghz (a Novelda X1) due to its ability to penetrate
soil. 

\subsection{Backscatter tag}

RF backscatter is the principle behind radar, but it has also long
been used as a low-power communication technique. RFID, for example,
uses an antenna as a reflector and changes the impedance to modulate
information on top of the reflected RF. The simplest kind of
modulation is binary on-off keying, which toggles the antenna between
grounded and open. 

By using backscatter to communicate, instead of an active radio chain,
these backscatter tags use orders of magnitude less power than
traditional radios. Since we are burying our tag underground, long
battery life is a primary design goal. Backscatter tags can be passive,
semi-passive or active. Passive tags, such as those in anti-theft
stickers or door security badges, both harvest their operating power
via RF and communicate using backscatter. This depends on having an
incoming source of RF with a strong enough signal to enable power
harvesting. Semi-passive tags such as~\cite{Zhang2017} use a battery
instead of harvesting RF power. Active tags are used in long-distance
scenarios, such as toll transponders. These tags both use a battery
\emph{and} amplify the outgoing backscatter reflection. They do not
have a full radio chain so they still rely on incoming RF to
communicate.

There are no off-the-shelf backscatter tags designed for radar, so we
built our own prototype (see Fig~\ref{fig:tag}). UWB radars transmit
at low power to avoid causing interference, so passive tags are not an
option because the incoming RF is far too quiet for power
harvesting. Instead, we implemented semi-passive and active
designs. The semi-passive tag has a very simple design, consisting of
a UWB Vivaldi antenna, an RF switch and an oscillator. A waterproof
case creates an air pocket around the antenna, which acts as a radome
and ensures proper impedance matching, as direct contact with soil could cause a mismatch. One might wonder why we need
anything beyond an antenna---is the strong reflection not enough? In
open air, the answer is yes. Underground, though, the tag is just one
reflector among many, many other reflective particles of dirt and
rock. We also want a way to isolate reflections that are coming from the
tag.

\subsubsection{Identifying the tag among dense reflectors}\label{findTag}

Recall that radars are used to measure speed as well as distance. Our
key insight behind how we isolate the signal from the tag leverages
the fact that the environment the tag lies within is very
stable. Roots grow and water seeps, but at slow speeds. If we make the
tag seem like it is moving quickly, the signal will stand out strongly
against an effectively stationary backdrop.

Let the impulse the radar transmits be represented by $p(t)$, and the
received signal by $r(t)$. Then, the digitized sample for the $n$th
range bin can be written as

\begin{equation}
r[n] = \alpha p(nT - \tau)
\end{equation}

where $T$ is the sampling period, $\alpha$ is the complex attenuation
and $d_k$ is the distance of the range bin in meters. Time of flight,
$\tau_k = 2d_k/c$, is the time for a radar impulse to travel to an
object and then reflect back again.

Above, for simplicity we have assumed that there is only one reflector
per range bin, but in reality there are multiple reflections in a
range bin since dirt is small and dense. Taking that into account, the
resulting sample will then become be the linear combination of all $k$
reflectors in the same bin:

\begin{equation}
r[n] = \sum\limits_{k=1}^K \alpha_k p(nT - \tau_k)  
\end{equation}

Because pulse-based radars transmit at a regular interval known as
the pulse repetition interval (PRI), they can be used to obtain the speed of moving objects. If
an object is moving at a constant speed of $v$ m/s, then every frame
the object's ToF changes by $2v\Delta/c$ where $\Delta$ is the PRI.

This change in the time-domain corresponds to a phase change in the
frequency domain: $\phi_k = 2\pi f (2v_k\Delta/c)$. Although the phase
changes for each frequency within the bandwidth of the impulse, we can
simplify the math by using only the radar's center frequency
$f_c$\footnote{This approximation is only valid for signals where the
  bandwidth of the signal is small compared to the center
  frequency. Some radars use pulse compression to help overcome this
  issue}. Then, the value for the $n$th bin in the $m$th frame will be

\begin{equation}
r_m[n] =  \sum\limits_{k=1}^K \alpha_k p(nT - \tau_k)e^{-j2\pi  \frac{2 v_k \Delta(m-1)}{\lambda}}  
\label{eq:timedomain}
\end{equation}

where $\lambda$ is the wavelength of the radar center frequency,
$f_c$.

Note how similar Eq.~\ref{eq:timedomain} is to the discrete Fourier
transform:

\begin{equation}
F_k = \sum\limits_{k=1}^{N-1} f_n e^{\frac{-j2\pi}{N}kn}
\end{equation}

If we apply a 1-D inverse Fourier transform to each range bin across a
collection of P pulses, we get a \emph{range-Doppler image} which
tells us the speed of moving objects:

\begin{equation}
R[n,s] = \frac{1}{P}\sum\limits_{i=1}^{P} r_m[n] e^{j2\pi\frac{(i-1)(s-1)}{P+1}}
\label{eq:rangedopp}
\end{equation}

In Eq.~\ref{eq:rangedopp} above, $s$ is the Doppler bin and $n$ is
the range bin and $P$ is the total number of pulses transmitted over
the collection time. 

\begin{figure}[h!]
  \centering
  \includegraphics[scale=0.4]{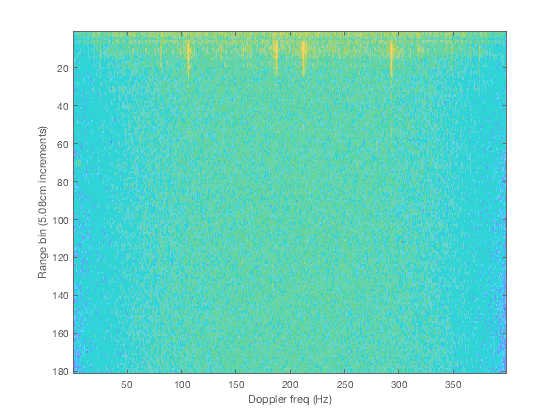}\\
  \caption{Range-Doppler image generated from a 10s radar capture. The bright spots at 212 and 293 Hz correspond to the oscillation of an antenna that is
grounded and ungrounded at 212 Hz. The additional frequencies are harmonics caused by the square-wave nature of the tag pattern. }
  \label{figure:rdplot}
\end{figure}

Figure~\ref{figure:rdplot} shows an example range-Doppler plot. We can
see that there are bright spots at 212 and 293 Hz. This plot is a 10
second capture where the radar is pointed at an antenna that is
grounded and ungrounded at 212 Hz.If there are no other
fast-moving objects in the radar's field of view, and the frame rate
of the radar is sufficiently high, we can use this property to discern
the signal from our backscatter tag from the signals due to other
reflectors.

Ultimately we set our tag to oscillate at 80Hz, which is below the Nyquist
frequency of our 200fps frame rate, but between the 60 and 120Hz
interference caused by AC power\footnote{this is only relevant for our
  indoor experiments}.

\begin{figure}[h!]
  \centering
  \includegraphics[scale=0.4]{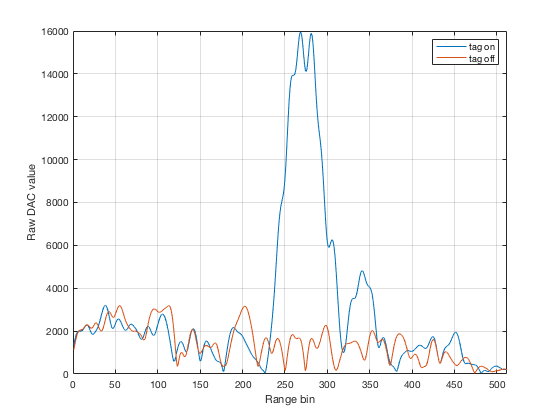}\\
  \caption{Plot of the range-Doppler vector corresponding to 80Hz, the
    frequency our tag oscillates at. When the tag is on, we see a very
    strong peak compared to when the tag is off}
  \label{fig:peak}
\end{figure}

Figure ~\ref{fig:peak} shows the plot of the vector corresponding to
the 80Hz frequency bin in our range-Doppler matrix. A strong peak
appears in bin 268 only when the tag is on. Thus we've successfully
identified the signal from the tag. This approach has the additional advantage the SNR of the signal of the tag increases with integration time. That is, the more frames we capture, the better the signals gets. So if a capture yields an ambiguous peak, we can simply capture additional frames. 

\subsubsection{Amplification}

A semi-passive tag works well in typical soil conditions, but for tags
in especially deep and/or wet soil, the tag signal is too weak to
reliably detect. Therefore we also designed an active variant of our
tag that adds an amplifier. The incoming RF is amplified before going
into the RF switch, which has another antenna attached to it. The
oscillator causes this second antenna to toggle on and off with the
amplified signal, allowing the tag to operate in more extreme
conditions. This active tag is completely indistinguishable to the
radar receiver, so we can use the same procedure to measure soil
moisture for both tags.

\subsection{Putting it all together}

This system relies on the user knowing a.) where the tag is located in
the field and b.) how deep the tag is buried. With this information we
know $d_s$, the amount of soil the direct path of RF has to travel
through (Fig.~\ref{figure:tagModel}). A few simple options make tracking this data easy, such as using marking flags,
annotated GPS coordinates, or slightly changing the oscillation
frequency of each tag allowing for a lookup table that maps between
frequency and tag depth/location. Taking measurements with the radar
not located directly over the tag does add error. Using directional
antennas in both the tag and radar helps minimize the chances of
taking measurements off-center, since the signal strength will be much
weaker when the radar isn't directly overhead. Another option is using
a more sophisticated radar with multiple receive antennas, which would
enable adjusting $d_s$ with angle of arrival data.

Another piece of information is still needed to measure soil moisture: the distance between the radar and the surface of the ground,
$d_a$. The radar itself can easily measure this. In our experiments, $d_a$ ranged between 0.5-2m. 

Then, using the technique outlined in~\ref{findTag}, our system finds the
range bin of the backscatter tag, $b_r$. The expected range bin if
there were no dirt on top of the tag is $b_T = d_a/r + d_s/r$, where
$r$ is the range resolution of the radar.

Now we can calculate $\Delta \tau$, the change in ToF caused by the
soil:

\begin{equation}
\Delta \tau = \frac{(b_r - b_T)r}{c}
\end{equation}

The approximate apparent dielectric constant of the soil is
\begin{equation}
  K_a \approx \left(\frac{c\Delta \tau}{d}\right)^2
\end{equation}

Finally, this apparent permittivity is fed directly into a known
equation like the Topp equation to determine VWC.

\section{Implementation}

\hspace*{-6em}
\begin{table*}
  \centering
  \caption{Always-on power consumption of prototypes \label{table:power_breakdown}}
  \begin{tabular}{l|l|l|l|l|l|l|l}
  & \textbf{Oscillator} & \textbf{RF switch} & \textbf{Power management} & \textbf{RF detector} & \textbf{Amplifier} & \textbf{MCU}  & \textit{\textbf{TOTAL}} \\ \hline
  \textbf{Active tag}  & 2.7uW               & 63uW               & 51uW                      & 87mW (3uW shutdown)  & 267mW              & 378uW (2.2uW) & 354.495mW               \\ \hline
  \textbf{Semi-passive tag} & 2.7uW               & 63uW               & 51uW                      & ---                  & ---                & ---           & 0.116mW                
\end{tabular}
\end{table*}

The radar chip we used was the X1 (NVA6100) by Novelda, which is
centered at 1.5Ghz and has a bandwidth of 3Ghz. The chip is \$100 per
unit. We interface with the radar via a development kit made by Flat
Earth Inc~\cite{chipotle} that runs on a BeagleBone Black single-board
computer. For these evaluations the radar captures were processed via
MATLAB, but the signal processing required could relatively easily be
ported to run in a low-level language on on a BeagleBone or
smartphone. All of our source code will be released to ensure
reproducibility.

\begin{figure}
    \includegraphics[width=55mm]{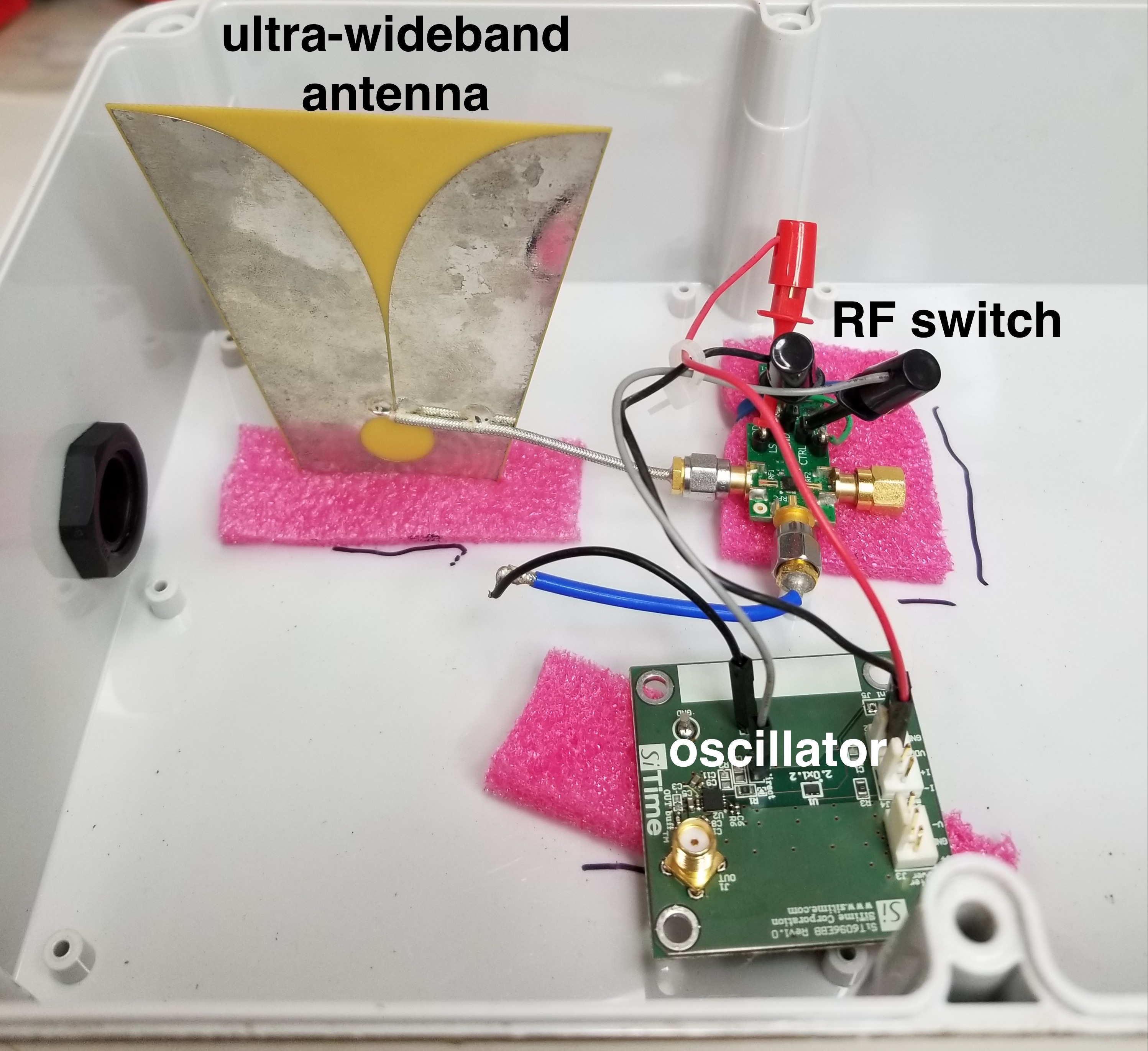}
\caption{Prototype of a semi-passive tag}
\label{fig:tag}
\end{figure}

The backscatter tags have three primary components: an SiT1534
programmable oscillator, an HMC1118 RF switch and a Vivaldi
ultra-wideband antenna (see Fig.~\ref{fig:tag}). A TPS76933 voltage regulator manages power
when the tag is powered by battery. The active tag has an additional antenna and an HMC374 amplifier.

\subsection{Power consumption}
The primary area of concern with regards to power consumption is the
backscatter tag, since it is underground and the batteries cannot be
easily replaced. The power consumption of the radar is still
important, especially if the readings will be collected via drone, but
we assume that the radar reader system can be charged at least
daily. The NVA6100 radar chip we use consumes 116mW of
power~\cite{X1datasheet}, and the entire reader system (radar chip
plus BeagleBone Black board) consumes 450mW, about a
quarter of what smartphones consume. Power consumption could be further reduced in the future by using a low-power microcontroller platform (e.g. MSP430) instead of a BeagleBone
Black.

The power consumption for both the active and semi-passive tags is
presented in Table~\ref{table:power_breakdown}.  The battery lifetime
of the semi-passive sensor is projected to be 15.02 years on
4$\times$AA batteries rated for 2500mAh. The active sensor consumes an
order of magnitude more power, so without duty cycling the battery
life would be about two months. However, using an RF detector such as the LT5538 that is powered up once per second to check for a wake signal, the battery
lifetime could be 3-4 years\footnote{assuming the high-power
  components wake for a total of 5-7 minutes per day}. This comes at
the cost of a more complicated system---the transmit power of UWB
radar is required to be very low by federal regulation in most
countries, which makes it insufficient for providing a wake
signal. However, since the tag antenna is wideband, a narrowband
signal such as WiFi or RFID can be used to wake up the tag
instead. Narrowband signals can be transmitted at powers up to 4W when
a directional antenna is used. Using a VNA we measured the attenuation of an omnidirectional wake signal centered at 2.4Ghz, and found that travelling through 30cm of fully-saturated clay loam causes losses of about 80-90dB. A 30-36dB transmission would be high enough power to overcome that and successfully activate many RF detectors.

Unless the tag needs to be deployed in adverse conditions where
the soil is extremely wet and/or has high clay content, the active sensor is
probably not worth the added
complication and decreased battery life.

\section{Evaluation}

We evaluate our system in both laboratory and \emph{in situ}
settings. The laboratory evaluations were done using a large bin
containing about a cubic meter of soil to ensure that the backscatter
tag is covered equally by soil on all sides. The \emph{in situ}
evaluations were done at a local organic farm whose fields contain
sandy silt loam soil (see Fig.~\ref{fig:farmSetup}). We used the same
local tap water for all experiments. Below we discuss experimental
considerations in more detail.

\subsection{Soil type selection}

\begin{figure}
    \includegraphics[width=65mm]{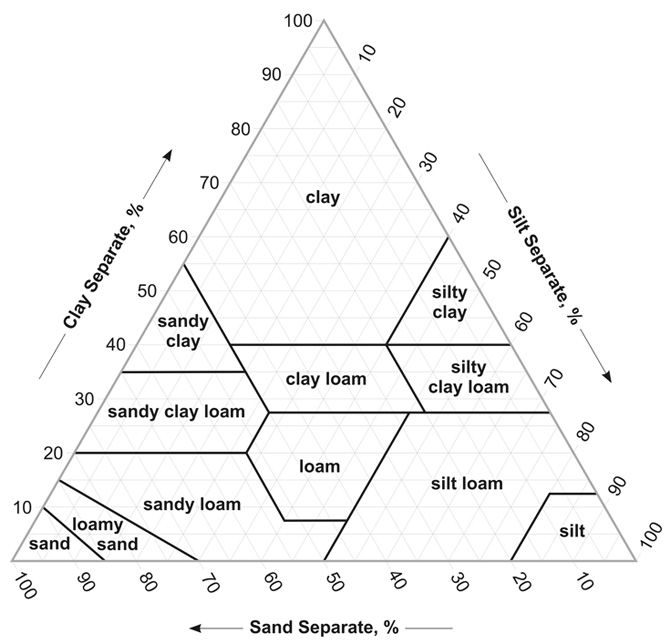}
\caption{Soil textural triangle~\cite{soilTriangle}}
\label{fig:soilTriangle}
\end{figure}

Soil can be broadly classified into three main types: clay, sand and
silt. Soils that are purely one type are rare, however, and most are a
mixture of two or more types (see Fig.~\ref{fig:soilTriangle}). Loam
is a roughly even combination of clay, silt and sand. Loamy soils are
considered to be ideal for agriculture~\cite{loamIdeal}, and most
crops are grown in soil that lies within the loam spectrum. We test
 on three sub-types of loam: sandy clay loam, silt loam and
clay loam. Testing a variety of soils is important because the soil
type can strongly impact RF propagation properties. For example, clay
soils have fine particle that allow it to hold water very well, and
they also tend to high higher organic matter content which impacts the
electrical conductivity (EC), which makes it more difficult for RF to
penetrate.

We collected our soils by consulting a recent map of farmland soil
classifications~\cite{soilMap}. All three soils are identified by the USDA as suitable for agriculture.

\subsection{Electrical conductivity}
As mentioned earlier, the depth that RF of a given wavelength can
penetrate into the ground (skin/penetration depth) is heavily impacted
by the soil's moisture content (which dictates the permittivity) and also
the EC.

Increasing water content increases both the permittivity and EC. Soil
with high organic matter content sees a greater increase in EC when
water is added. For example, our measurements found that the EC of
fully-saturated potting soil is 10x that of fully-saturated sandy clay
loam. It still works in potting soil, but the maximum reliable deployment
depth is only 10-15cm vs the up to 75cm possible with loam soils typically seen on
farms.

Furthermore, soil amendments such as compost or liquid fertilizer can
also increase the EC of soil. Most farms maintain an EC between
0.75-2mS/cm~\cite{idealEC}. In our experiments the EC remained
within that range for all soils except potting soil. Fortunately, we
did not see significant RF attenuation until EC levels rose above
2mS/cm.

\subsection{Root zone depth}
Root zone depth, or maximum root zone depth, is the maximum depth of a
plant's roots. The effective root zone depth is the depth of soil that
a plant's roots extract the most moisture. About 70\% of the moisture
extracted by a plant's roots is from the top half of the maximum root
zone. For example, celery has a maximum root depth of 60cm, and
an effective root zone depth of 30cm. This means that moisture
should be monitored within the top foot of soil.

Most crops have an effective root depth between 15-60cm, however fruit
crops (especially those that grow on trees) can extend as deep as
75cm~\cite{rootZone}. Our laboratory experiments were done at
a depth of 30cm primarily due to the limitations of container sizes
that could ensure the sensor was covered on all sides, but our \emph{in situ} experiments (see Fig.~\ref{fig:SNRdepth}) suggest that it could be deployed at depths up 75cm.

\subsection{Calibration}
All soil moisture sensors require a one-time soil-specific calibration
to achieve high accuracy. One common calibration procedure is
gravimetric, which involves weighing wet samples, oven drying them to
calculate the ground-truth VWC, and then fitting those measurements to
the sensor readings that were taken at the time of sample collections
to produce a custom equation that relates sensor output to VWC.  We
performed gravimetric calibrations for both our commercial sensor and
radar sensor.

If lower accuracy is acceptable, soil-specific calibration is not
necessary and a general equation like the Topp equation can be used
instead.

Unlike other RF-based solutions such
as~\cite{Ding2019}~\cite{Aroca2018}, our system does not require any
additional calibration as compared to commercial soil
sensors. However, accurate records of the depth the sensor was
deployed at are required. This depth can be measured manually, or by
using the radar itself once the sensor is placed in the hole (but
before the soil is replaced). We use the latter approach in our
evaluations.

\section{Results}
\begin{figure}
    \includegraphics[width=55mm]{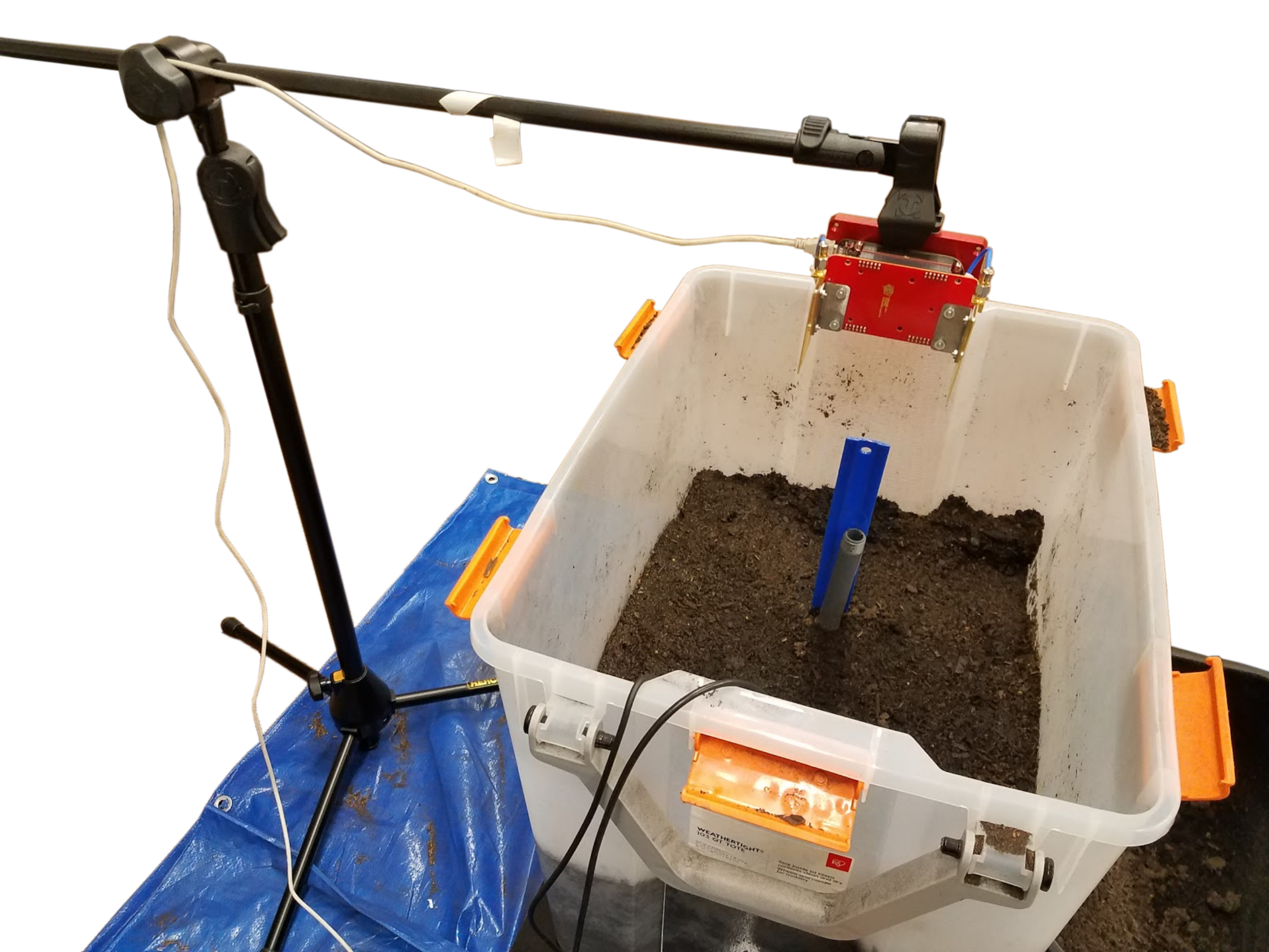}
    \caption{Setup for laboratory experiments}
\label{fig:farmSetup}
\end{figure}
\subsection{Laboratory}

\begin{figure*}
  \begin{minipage}[b]{0.55\textwidth}
    \centering
    \includegraphics[width=85mm]{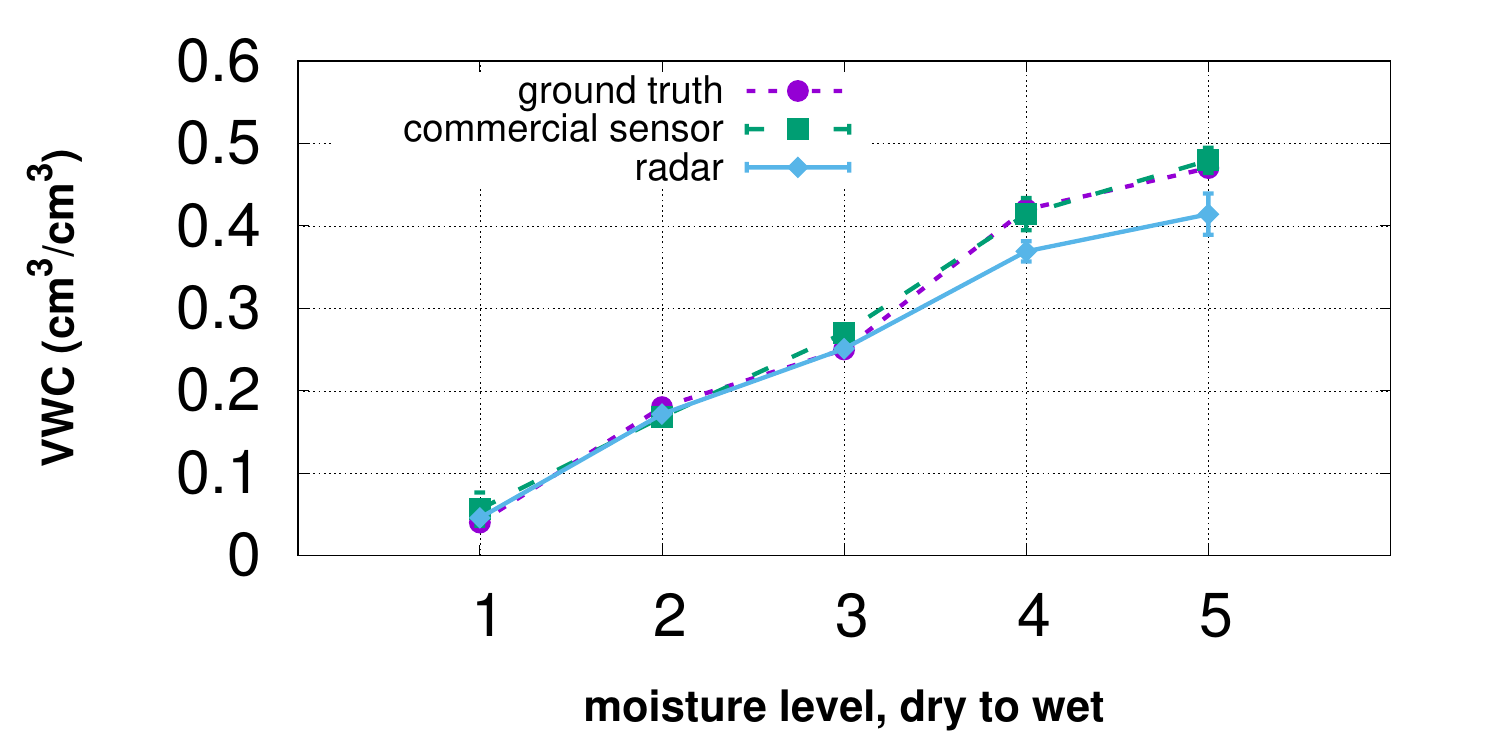}\\
    \subcaption{Sandy clay loam}
  \end{minipage}\\
  \begin{minipage}[b]{0.55\textwidth}
    \centering
    \includegraphics[width=85mm]{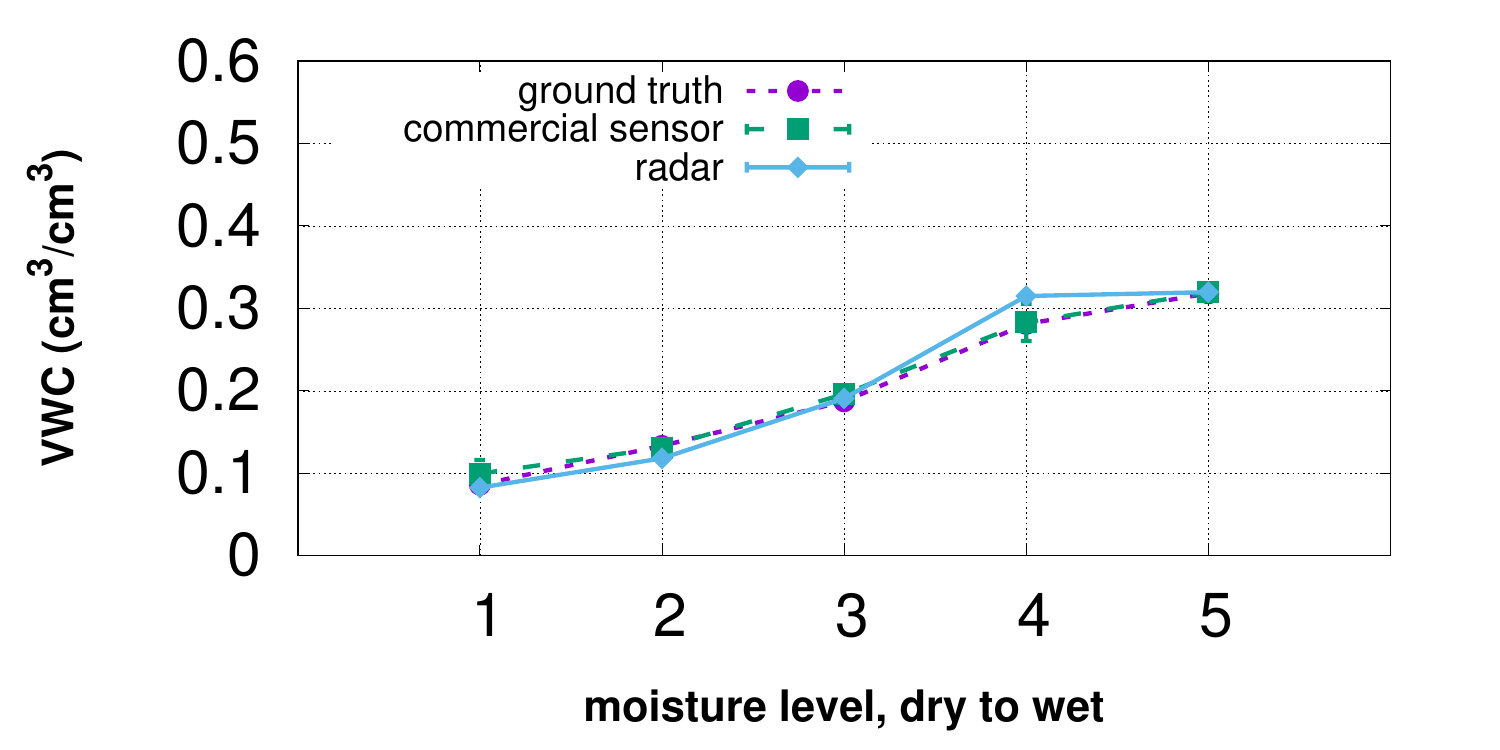}\\
    \subcaption{Silt loam}
  \end{minipage}\\
  \begin{minipage}[b]{0.55\textwidth}
    \centering
    \includegraphics[width=85mm]{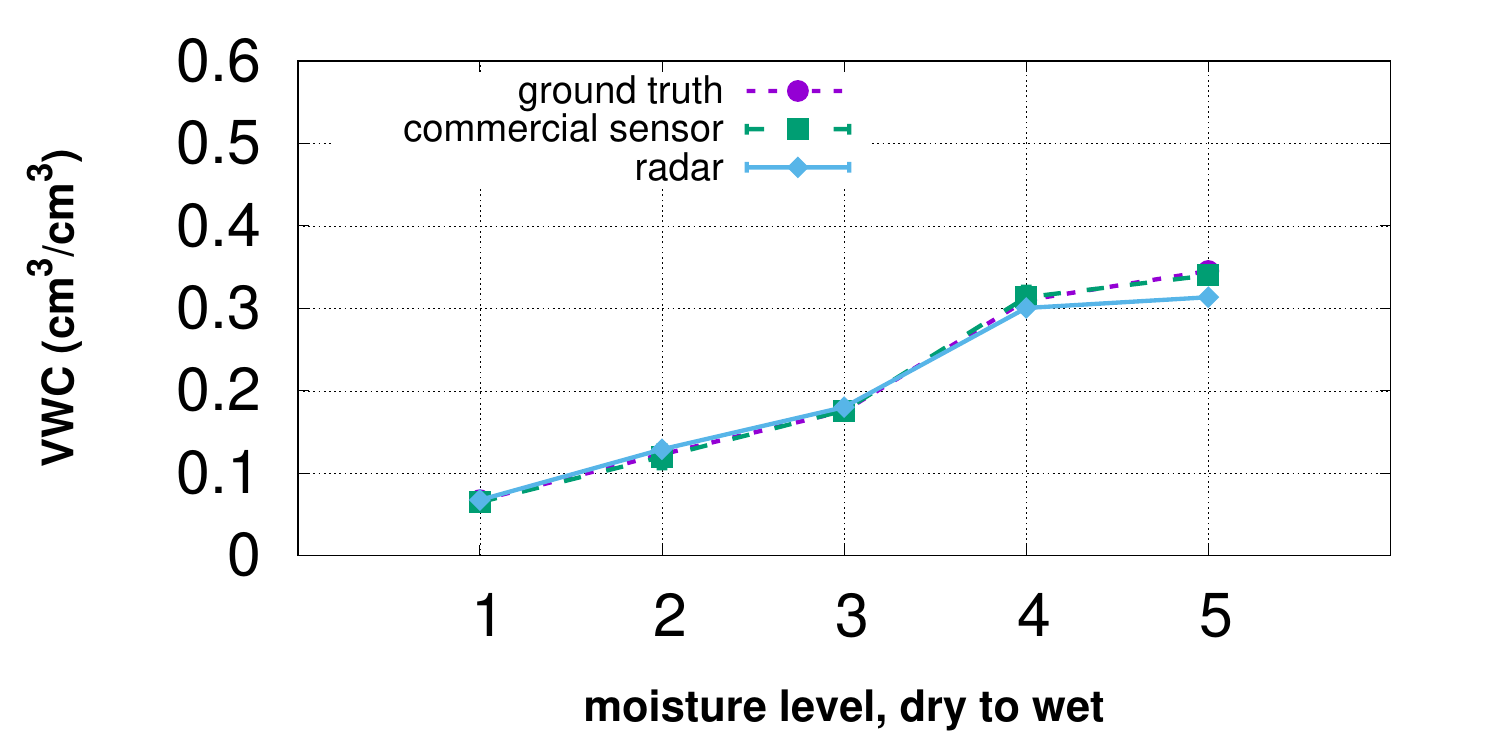}\\
    \subcaption{Clay loam}
  \end{minipage}%
  \caption{Active tag VWC measurements of three soil types from dry
    to saturated with the tag buried at a depth of 30cm. Moisture level 1 is completely dry soil which was then gradually dampened in 7 liter increments until saturation at level 5. Note that saturation depends on the soil type.}
\label{fig:active}
\end{figure*}

Figure~\ref{fig:active} shows the results from our active tag. Each
radar datapoint is the average and standard deviation of 10
measurements. The radar captures used for the measurements lasted
10-30s with the exception of the saturation moisture levels which used
100s\footnote{Faster measurements are possible with higher radar frame rates. The radar development kit we used runs a Linux distribution, and IO interrupts limited our achievable frame rate to 200fps. Porting the development kit to run on a barebones system may further increase the framerate without having to upgrade the radar hardware itself.} captures. Each commercial sensor datapoint is the average and
standard deviation of 3-5 measurements, where each measurement is
taken in a different part of the soil. The size of the container we
conducted experiments in limited the number of commercial sensor
datapoints. To conduct the experiment, we began with about a cubic
meter air-dried soil and gradually dampened in 7 liter
increments. In these laboratory experiments we homogenized the soil
moisture by mixing the added water vigorously by hand. This was to
ensure that the Teros 12 sensor we compared against was not biased by
wet or dry pockets of soil.

We see that for all soil types both our system and the commercial sensor
closely track the ground truth, which is the average of two oven-based
volumetric measurements per moisture level. The average error of
our system is $0.015cm^3/cm^3$, compared to $0.007cm^3/cm^3$ on the commercial Teros 12 sensor. Though our system'ss
average error is higher than the commercial sensor, it is not
significant. Calibrated commercial sensors are advertised having an
average error between $0.01-0.03cm^3/cm^3$. The greatest error is seen
with the sandy clay loam soil at saturation, where our system
underestimates VWC by $0.05cm^3/cm^3$. This maximum level of error is
also typical among commercial sensors.

\begin{figure*}
  \begin{minipage}[b]{0.55\textwidth}
    \centering
    \includegraphics[width=85mm]{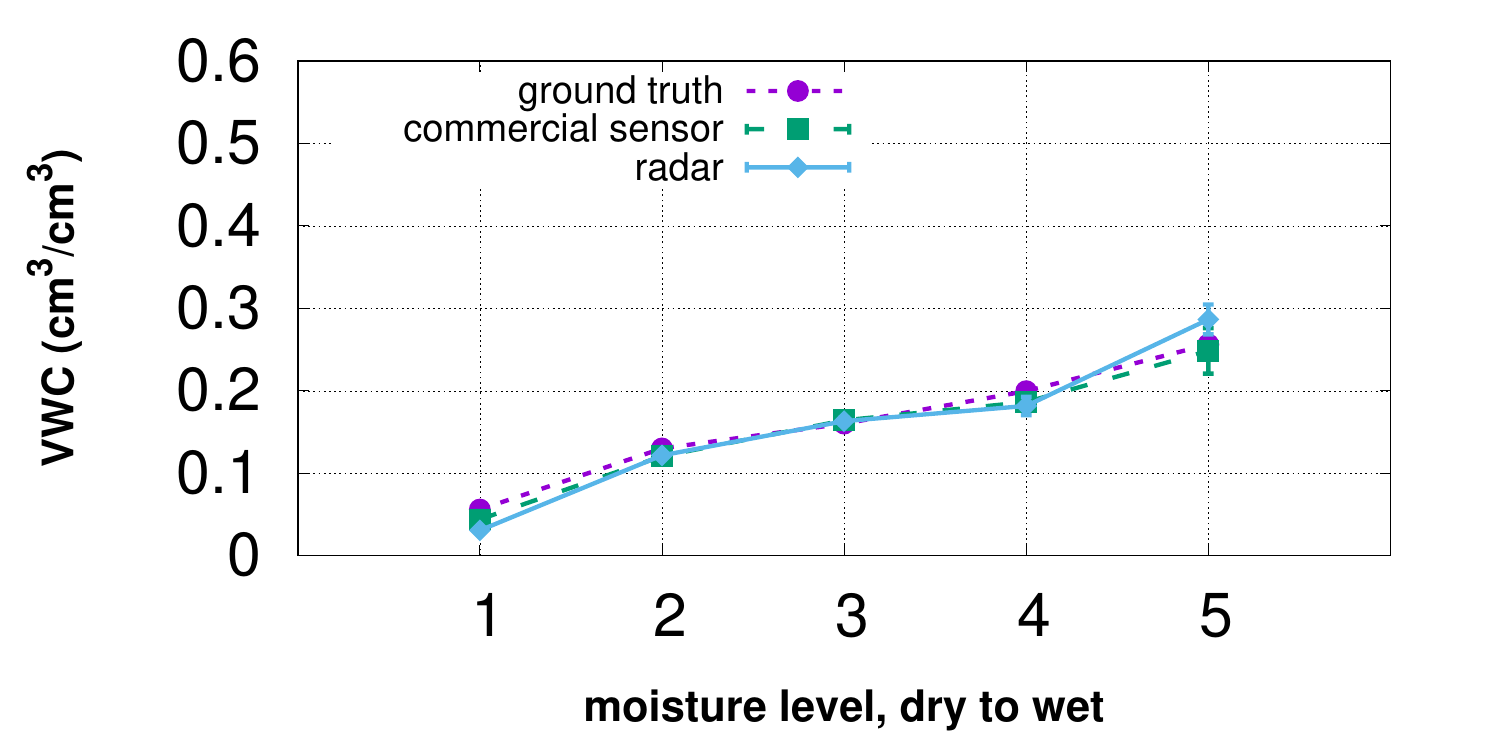}\\
    \subcaption{Sandy clay loam}
  \end{minipage}\\
  \begin{minipage}[b]{0.55\textwidth}
    \centering
    \includegraphics[width=85mm]{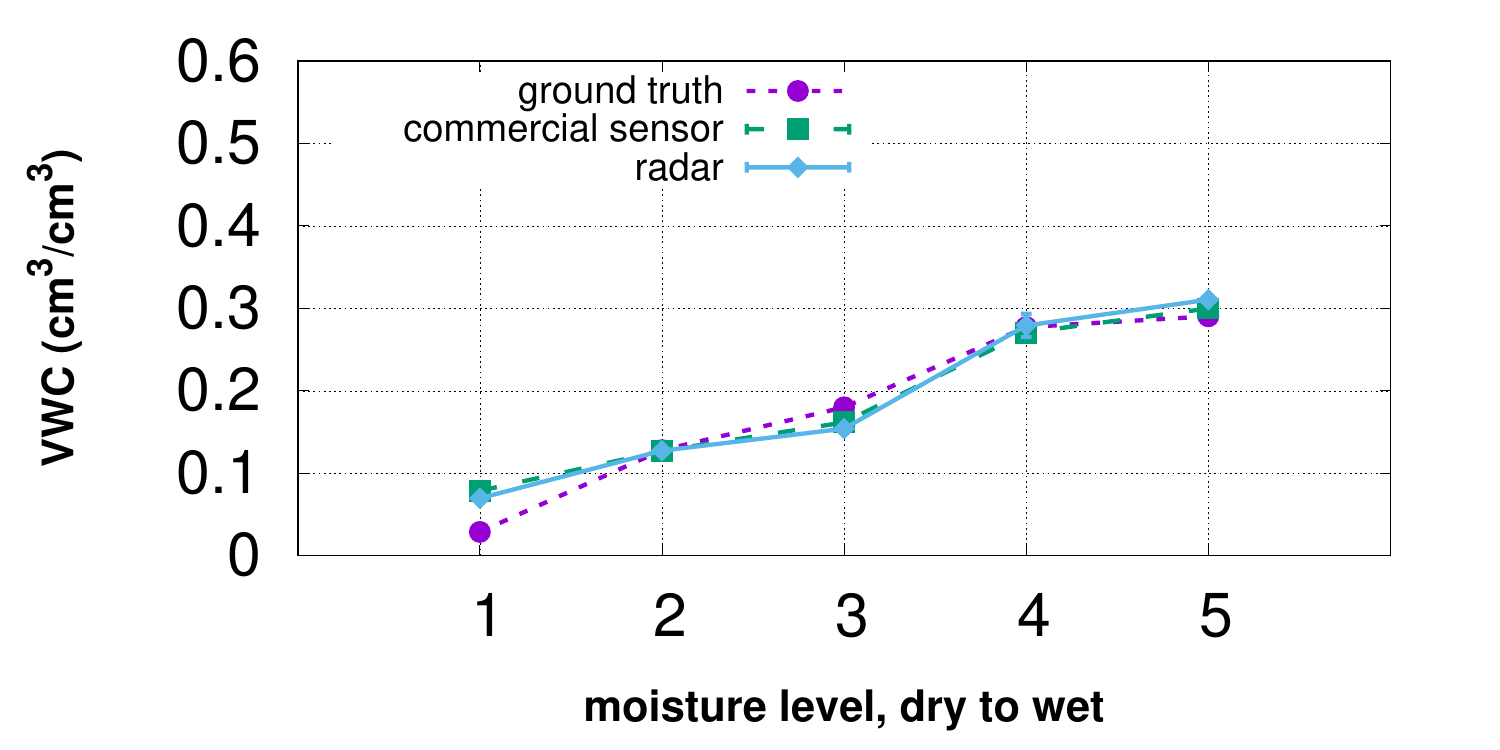}\\
    \subcaption{Silt loam}
  \end{minipage}\\
  \begin{minipage}[b]{0.55\textwidth}
    \centering
    \includegraphics[width=85mm]{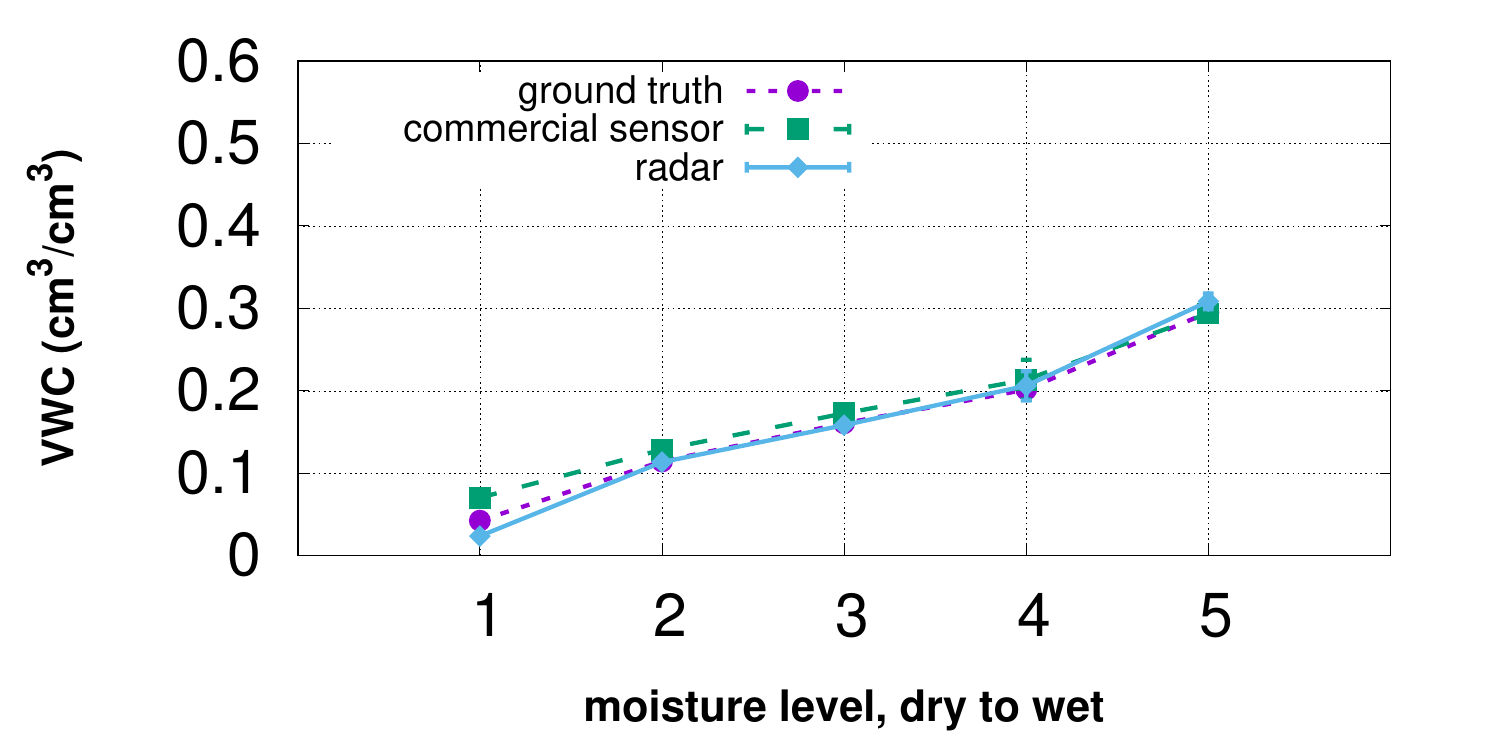}\\
    \subcaption{Clay loam}
  \end{minipage}
  \caption{Passive tag VWC measurements of three soil types from dry
    to loss of signal with the tag at a depth of 30cm. Moisture level 1 is completely dry soil which was then gradually dampened in 4-5 liter increments.}
\label{fig:passive}
\end{figure*}

Figure~\ref{fig:active} shows the results from our passive tag. This
time we added water in $4-5L$ increments and stopped when the signal
from the passive tag was no longer visible. This was typically 5-15\%
before saturation. Again, both our system  and the commercial sensor
closely track the ground truth, with our system  achieving an average
error of $0.014cm^3/cm^3$ and the commercial sensor
$0.013cm^3/cm^3$. The maximum error for both our system and the
commercial sensor was $0.04cm^3/cm^3$.

\begin{figure}
    \subcaptionbox{Active tag}{%
    \includegraphics[width=85mm]{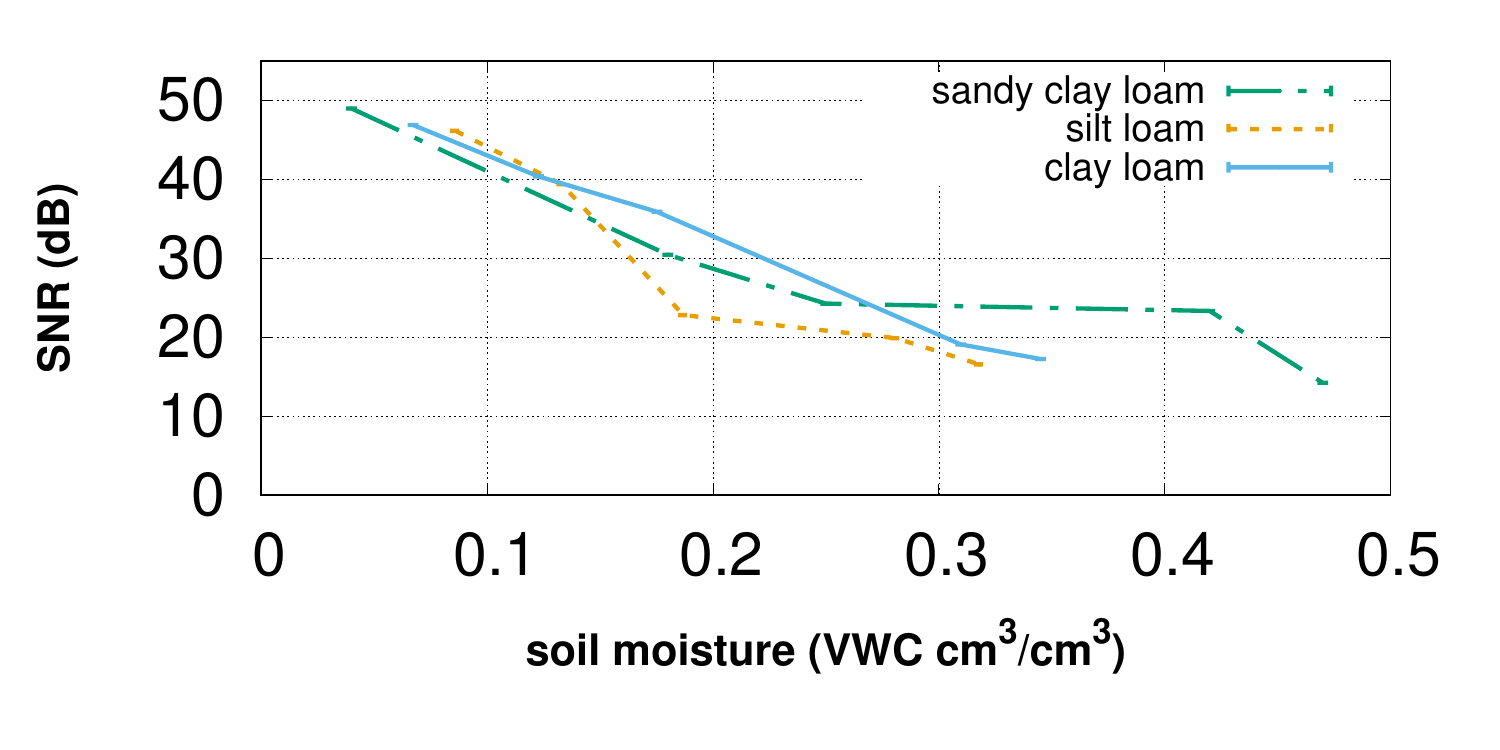}%
   } 
    \subcaptionbox{Passive tag}{%
    \includegraphics[width=85mm]{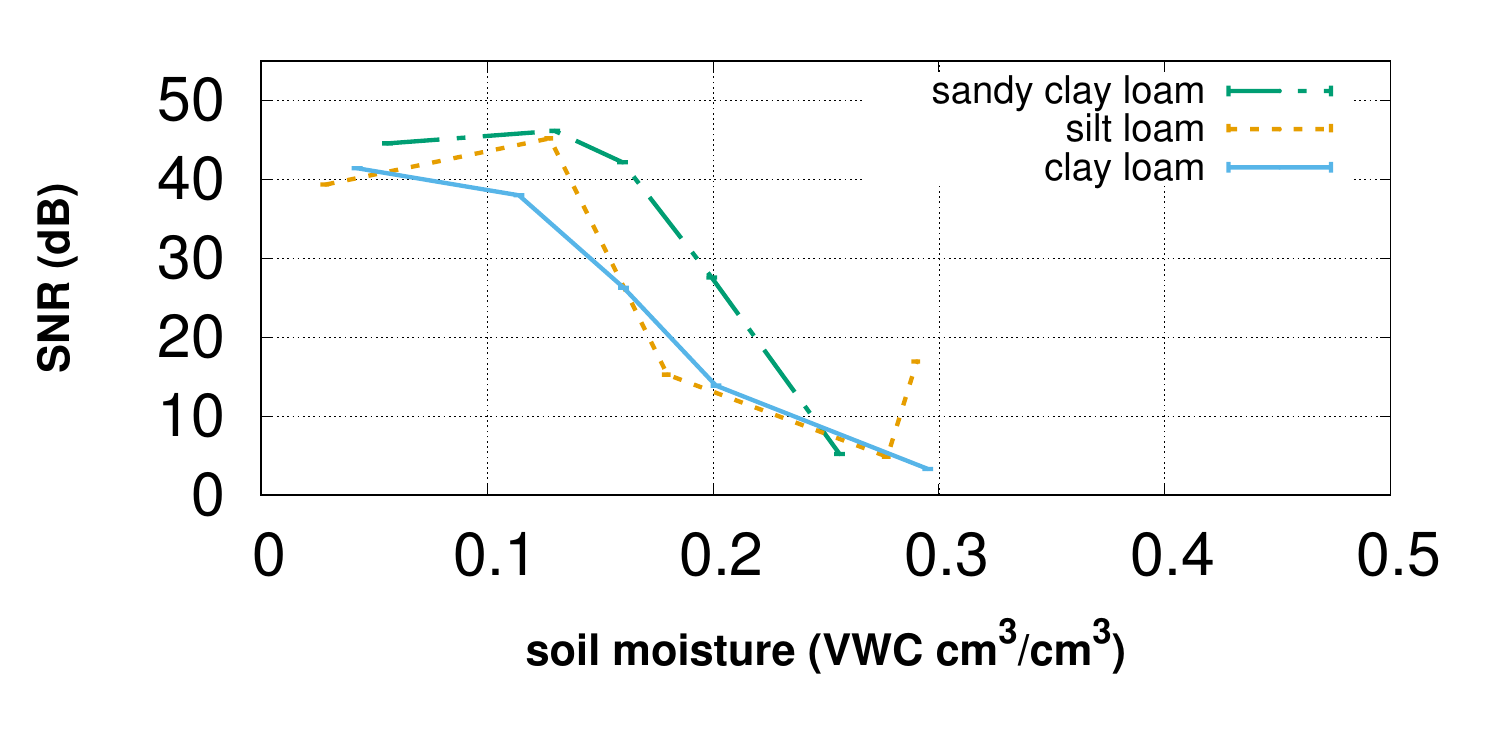}%
   }
   \caption{SNR vs VWC of 100s captures for passive and active tags in
     the laboratory}
\label{fig:snrLab}
\end{figure}

Figure~\ref{fig:snrLab} shows the results of SNR vs soil moisture for
both tags across the three different soil types. As expected, the SNR
for the passive tag decreases more quickly than the active tag as
moisture level increases. Also, the signal in both silt and clay loam
soils is weaker than the sandy clay loam. There does not appear to be
a significant difference between silt and clay loams, though.

\subsection{\emph{In situ}}
\begin{figure}
    \includegraphics[width=55mm]{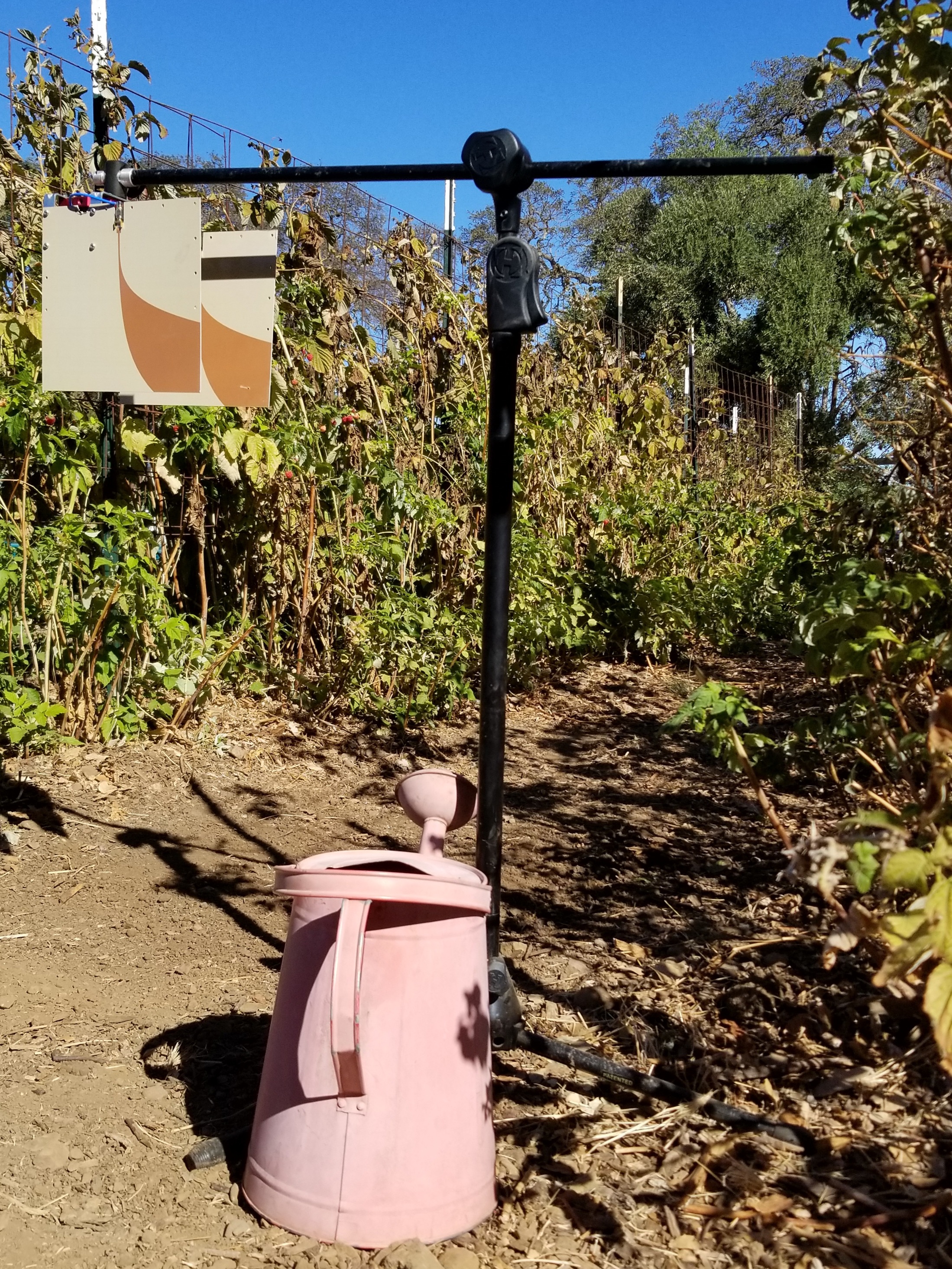}
\caption{Setup for \emph{in situ} experiments at a local farm. The
  radar is pointed at an underground tag, and the surface of the soil
  is watered in 7L increments using a watering can.}
\label{fig:farmSetup}
\end{figure}

\begin{figure}
  \subcaptionbox{Active tag}{%
    \includegraphics[width=85mm]{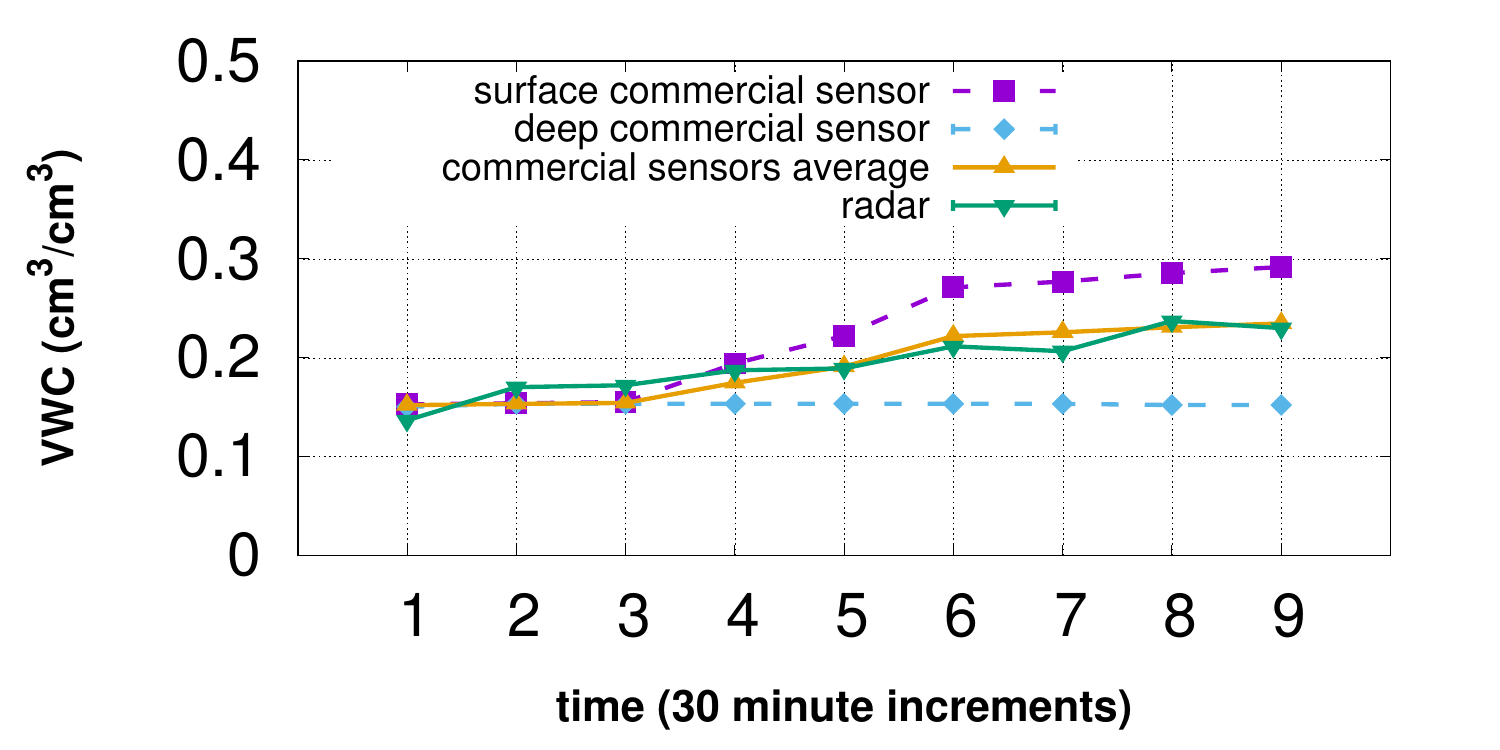}%
  }
    \subcaptionbox{Passive tag}{%
      \includegraphics[width=85mm]{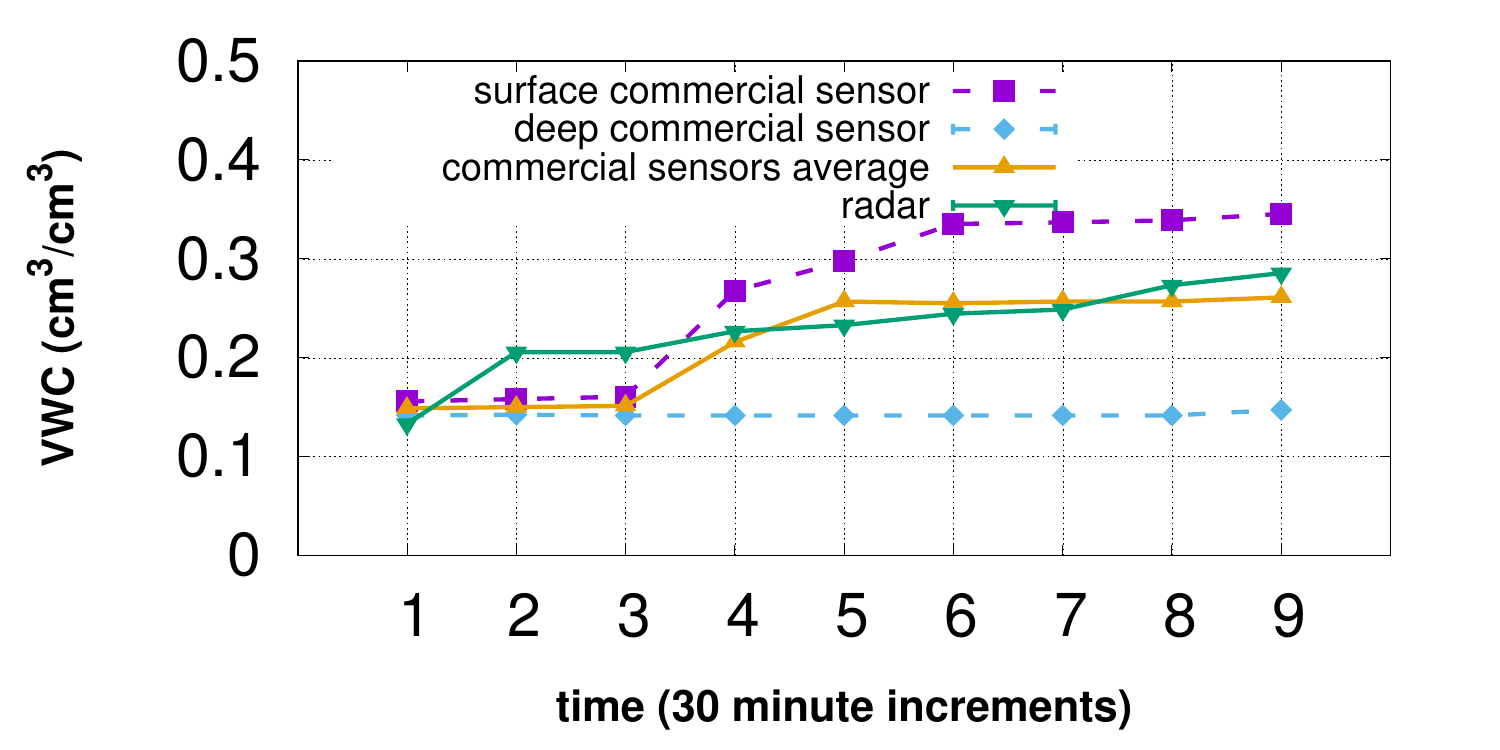}%
    }
      \caption{VWC \emph{in situ} on a farm field with passive and active
        tags buried at a depth of 30cm. Measurements were taken every
        30 minutes, with 7 liters of water poured on the soil at times 2, 4, 6
        and 8.}
        \label{fig:VWCsitu}
\end{figure}

Figure~\ref{fig:VWCsitu} shows the results of both passive and active
tags for the \emph{in situ} VWC experiments. In these experiments our
tag was buried under 30cm of soil. For comparison, we used two Teros
12 sensors, one at a depth of 30cm and the other near the surface at a
depth of 5cm. Unlike the laboratory experiments, we do not disturb the
soil and instead let the water seep over time. The deeper commercial
sensor showed no change in soil moisture across the experiments, even
after applying more than 20 liters of water and letting it seep
overnight. Water did successfully seep into the soil around the
shallow sensor, but there was still a delay of up to an hour between
watering and seeing the change of moisture level.

As expected, since our system  measures the average moisture of the soil
between the tag and the surface, it closely tracks the average of the two Teros
sensors. Furthermore, our system reacts immediately to the addition of
water. This suggests that it  provides faster feedback after water application
than traditional sensors, which might prevent
over-watering. Furthermore, it inherently reflects the average
soil moisture across the whole effective root zone, whereas multiple
commercial sensors are required to accomplish the same.

\begin{figure}
    \includegraphics[width=85mm]{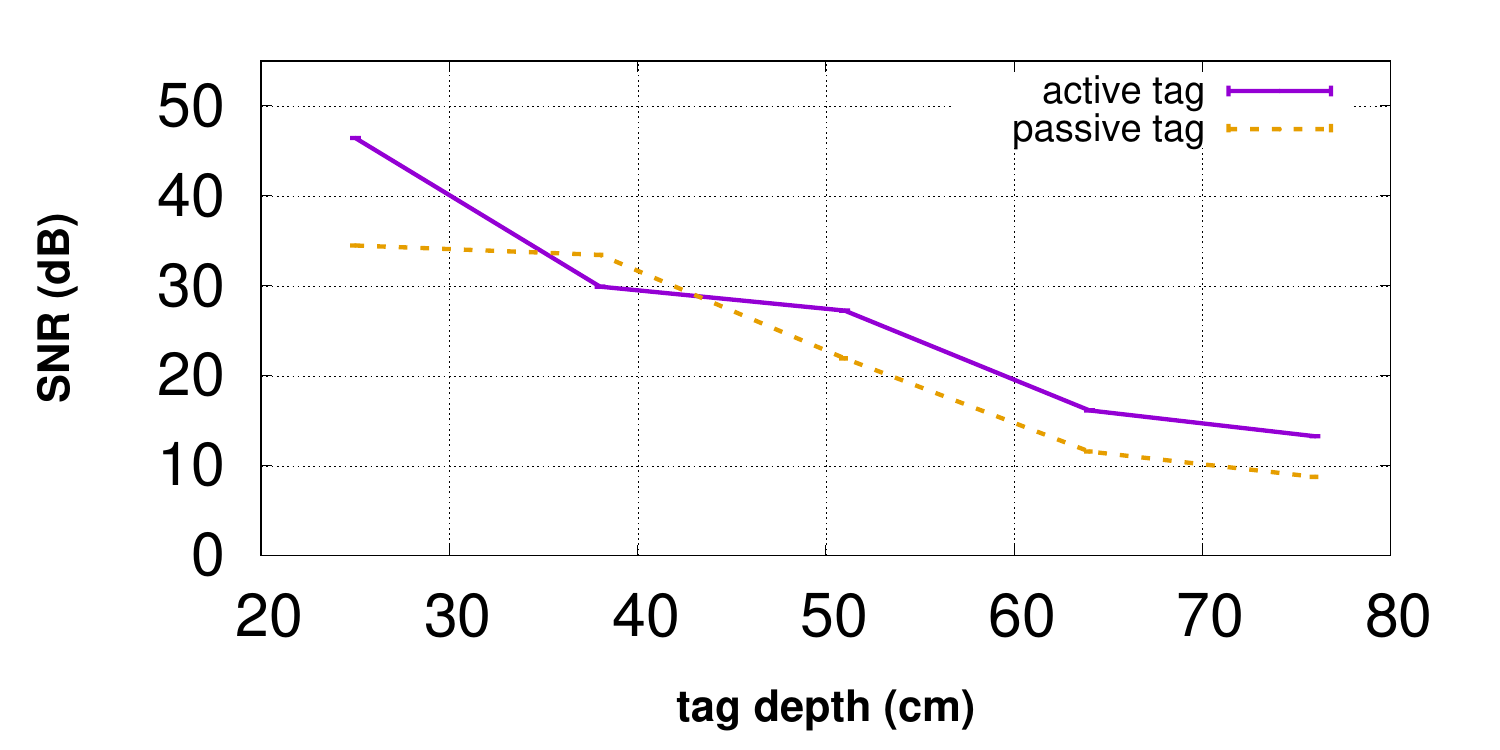}
\caption{SNR vs tag depth of 100s captures for passive and active
  tags. Measurements were performed in an actively watered farm field
  containing sandy clay loam. The VWC of the soil was about 15\% at
  the time of measurement. }
\label{fig:SNRdepth}
\end{figure}

One of the limitations of our laboratory experiments is that it is
difficult to evaluate with the tag buried more than 30 cm
deep, since covering the tag with $>$30cm of dirt on all sides would
require bringing a prohibitively large amount of dirt
indoors. Outdoors, we were able to dig a hole more than 75cm deep and
gradually cover the tag with dirt. Fig.~\ref{fig:SNRdepth} shows the
results of these experiments. At a VWC of $0.15cm^3/cm^3$ we were
still able to detect both tags successfully at a depth of 77cm. This
suggests that it can probably be deployed deeper than the 30cm
evaluated in-laboratory to accurately measure the VWC typically seen
on farms.

\begin{figure}
    \includegraphics[width=85mm]{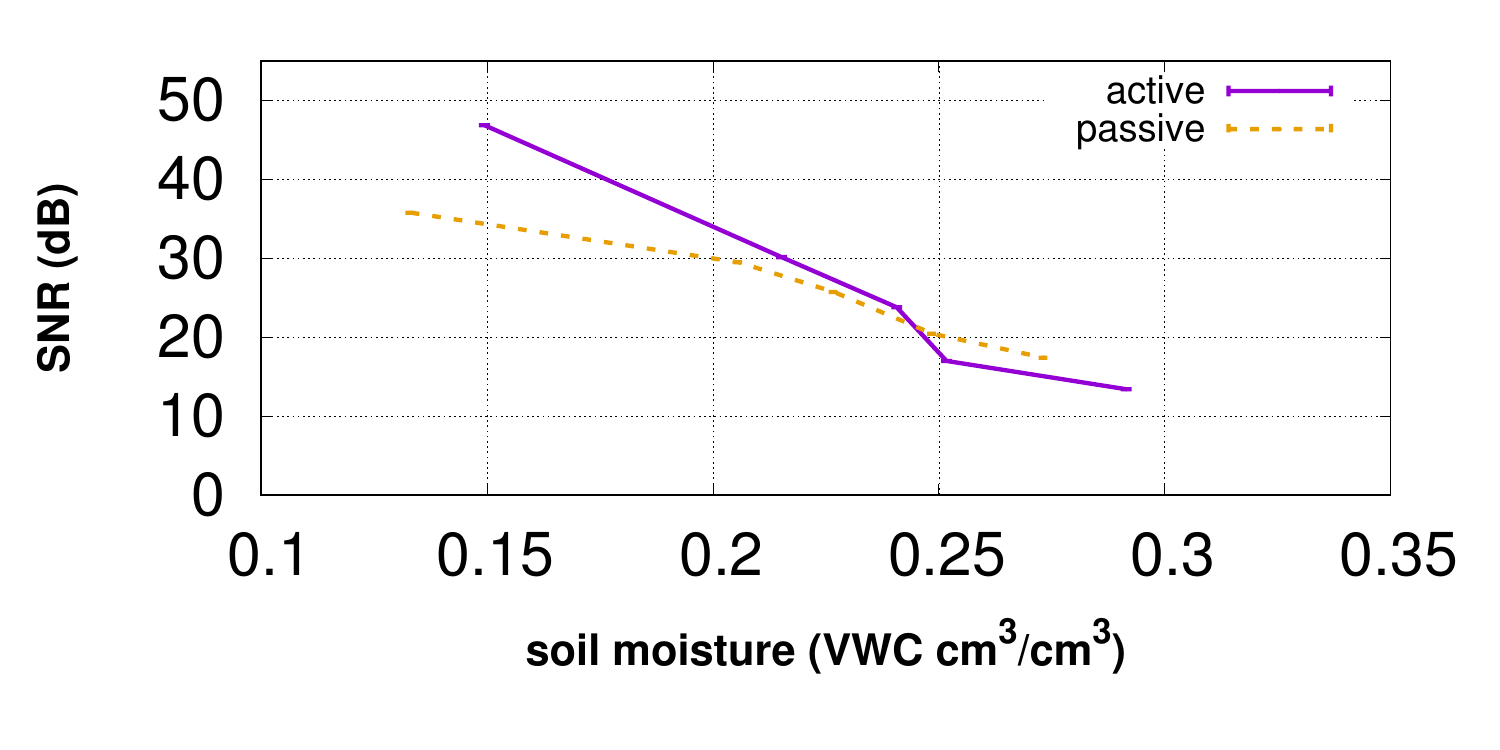}
    \caption{SNR vs VWC \emph{in situ} for 100s captures on a farm
      field with passive and active tags buried at a depth of 30cm}
\label{fig:SNRInSitu}
\end{figure}

Figure~\ref{fig:SNRInSitu} shows how the SNR for both active and
passive tags changes with VWC. Compared to the laboratory experiments
(Fig~\ref{fig:snrLab}), the passive tag SNR drops off much less
steeply. This further suggests that the passive version of the tag is
well-suited for agriculture and that the added complication and
reduced battery life of the active tag will usually not be necessary.

\section{Related Work}
In addition to GPR techniques, other works have also used RF to
measure soil moisture. Strobe~\cite{Ding2019} uses commodity WiFi
transmissions to measure relative ToF between buried antennas. These
antennas require being wired to a power-hungry 802.11 WiFi chip. Furthermore, there are significant additional calibration procedures required. Non-radar UWB transceivers have also used ToF to perform sensing such as localizaton~\cite{grobetawindhager2019snaploc} and ECG~\cite{Toll2019WirelessEB}.

Researchers in Israel used ToF measured via a \$20,000 GPR paired
with buried metal bars~\cite{Shamir2018} to measure soil moisture. The
radar used requires direct contact with the surface of the soil,
though, and the accuracy to our system is similar.

Our system is inspired by RFID, which has itself been used to classify
food and beverages~\cite{Ha2018}~\cite{Wang2017} and measure soil
moisture~\cite{Aroca2018}. The latter work uses commodity RFID tags
paired with neural networks to determine soil moisture via RSSI. This
work has the drawback that the neural network has to be re-trained for
every single soil type \emph{and} deployment depth. However, they have
successfully gotten an agricultural robot to collect the moisture
measurements autonomously.

\section{Discussion and Conclusion}
We believe our system has the potential to be useful to farmers with both large and small farms. There are a number of steps between this work and
real-world deployment, however.

The environmental impact of the tags needs to be considered. Wireless
tags are more easily lost. Research into biodegradable printed circuit
boards~\cite{Guna2016} and soil batteries~\cite{Lin2015} may allow for
a more eco-friendly version that doesn't leach harmful elements into the soil over time. 

We acknowledge that there is a lot of engineering and political work
to go from laboratory prototype to field trials, especially trials in
developing nations. We hope to trial our system at local farms soon, and
someday at farms in developing nations.

We also realize that there is a lot of active research in how to
best utilize agricultural robots and drones. These technologies have
seen some adoption, but in general much research remains for
determining how to scale.

There are also a number of additional research opportunities we would
like to explore: 
\begin{enumerate}
    \item Creating a tag with two antennas/oscillators would allow us to relative ToF in addition to absolute. Our system currently measures absolute ToF, which corresponds to average soil moisture, but sometimes point soil moisture is preferred. Furthermore it can be calculated without knowledge of the tag deployment depth if the separation between antennas is known.
\item Encoding additional information into radar backscatter so that the tag itself can store information like deployment depth and location, eliminating the need for the operator to use a lookup system.
\item Sensing opportunities beyond soil moisture such as measuring EC and contaminant mapping.
\end{enumerate}

\section{Summary}

In this paper we presented a two-part system for sensing soil moisture with RF that combines low-cost backscatter tags with a consumer-grade UWB radar acting as a reader. We achieve completely wireless soil moisture sensing with an accuracy comparable to that of state-of-the-art commercial and scientific sensors at an order of magnitude lower cost. We acknowledge
that there is a large gap between small-scale prototype and systems
deployed at scale, but we believe that it has the potential to
become an effective soil sensing solution for farmers in both
developing and developed nations.

\bibliographystyle{ACM-Reference-Format}
\bibliography{ipsn}

\end{document}